\documentclass[]{aa}
\usepackage{natbib}
\usepackage{graphicx}
\usepackage{color}
\usepackage{txfonts}

\bibpunct{(}{)}{;}{a}{}{,}

\begin{document}

\title{On the onset of galactic winds in quiescent star forming galaxies}

\author{Yohan Dubois \inst{1} and Romain Teyssier \inst{1}}

\offprints{Y. Dubois}

\institute{Service d'Astrophysique,
CEA/DSM/DAPNIA/SAp, Centre d'\'Etudes de Saclay, L'orme des Merisiers,
91191 Gif--sur--Yvette Cedex, France
 \\
\email{ydubois@cea.fr}}
\date{Accepted; Received; in original form;}

\label{firstpage}

\abstract  {The hierarchical  model of  galaxy formation,  despite its
many  successes, still  overpredicts  the baryons  fraction locked  in
galaxies as a condensed  phase. The efficiency of supernovae feedback,
proposed a  long time ago as  a possible solution  for this so--called
``overcooling''  problem,  is   still  under  debate,  mainly  because
modelling  supernovae  explosions   within  a  turbulent  interstellar
medium, while capturing realistic  large scale flows around the galaxy
is a  very demanding task.   } {  Our goal is  to study the  effect of
supernovae feedback on  a disk galaxy, taking into  account the impact
of  infalling  gas  on  both   the  star  formation  history  and  the
corresponding outflow structure, the apparition of a supernovae-driven
wind being highly sensitive to the  halo mass, the galaxy spin and the
star formation efficiency.  }  {  We model our galaxies as cooling and
collapsing NFW  spheres.  The dark  matter component is modelled  as a
static external potential, while  the baryon component is described by
the  Euler  equations  using  the AMR  code  RAMSES.   Metal-dependent
cooling   and   supernovae-heating    are   also   implemented   using
state-of-the-art  recipes coming  from  cosmological simulations.   We
allow for 3  parameters to vary: the halo  circular velocity, the spin
parameter and the star formation efficiency. } { We found that the ram
pressure  of  infalling  material  is  the  key  factor  limiting  the
apparition  of   galactic  winds.  We  obtain  a   very  low  feedback
efficiency, with  supernovae to  wind energy conversion  factor around
one percent, so that only  low circular velocity galaxies give rise to
strong winds. For massive galaxies,  we obtain a galatic fountain, for
which we  discuss the observational  properties.}  { We  conclude that
for quiescent  isolated galaxies, galactic  winds appear only  in very
low mass systems.  Although that can quite efficiently  enrich the IGM
with metals, they  don't carry away enough cold  material to solve the
overcooling problem. }

\keywords{galaxies: formation -- galaxies: evolution -- galaxies: structure -- methods: numerical }

\authorrunning{Dubois,   Teyssier} 
 
\titlerunning{On the onset of galactic winds in quiescent star forming
galaxies}

\maketitle

\section{Introduction}

The  hierarchical  model  of  structure  formation,  for  which  massive galaxies grow by mergers of smaller satellites  or by
filamentary accretion of  gas, is now a  well-established theory (the
so--called CDM paradigm),  that compares favorabily with observations.
In  this standard  picture, dark  matter  is the  main component  that
drives   structure  formation   via  gravitational   instability  from
primordial  density  perturbations.   On  the  other  hand,  numerical
simulations  including gas  dynamics with  radiative cooling  and some
recipes of star formation tend  to produce galactic disks that are too
small, to dense, with too  low angular momentum and too many condensed
baryons, when compared to observations. These various problems, called
``overcooling'' (\citealp{dekel&silk86, blanchardetal92, navarro&white93, yepesetal97, gnedin98, hultman&pharasyn99, somerville&primack99, coleetal00, kayetal02, springel&hernquist03}) or ``angular momentum'' 
(\citealp{navarro&benz91, sommer-larsenetal99, steinmetz&navarro99, bullocketal01, maller&dekel02, abadietal03, sommer-larsenetal03, read&gilmore05}) problem, may all have the same
physical origin.  Nevertheless, they are considered as  weak points in
the CDM theory.

Supernovae--driven  winds  are  a  key ingredient  of  current  galaxy
formation  models, in  order to  suppress the  formation  of low--mass
galaxies  and maybe to  solve the  ``overcooling'' problem.   They can
potentially suppress the fast  consumption of gas by transferring cold
and  dense  gas into  the  hot and  diffuse  surrounding  medium of  a
galaxy. Thus,  a fraction  of gas would  not collapse into  cold dense
molecular clouds and would not  transform into stars.  That is the most
favored explanation for the disruption  of star formation and the lack
of   baryons   available   in  stars   (\citealp{springel&hernquist03,
rasera&teyssier06,  stinsonetal06}).   Feedback  is  also  invoked  to
reproduce   the   large   disc   galaxies  observed   at   our   epoch
(\citealp{weiletal98, thacker&couchman01}), the morphology of galactic
discs (\citealp{sommer-larsenetal03, okamotoetal05}), to stabilize the
gaseous       disc      against      the       Toomre      instability
(\citealp{robertsonetal04})   or  to  fit   the  temperature   of  the
intergalactic     medium     (IGM)     at    redshift     $\sim$     3
(\citealp{cen&bryan01}).

The proper modelling  of galactic winds is quite  difficult to handle,
especially  in cosmological  simulations,  because of  the very  large
scale  separation between  the galactic  outflow  in the  IGM and  the
supernova  explosions within  the interstellar  medium (ISM).   But it
represents an unavoidable task in order to solve the weakest points of
the    hierarchical   model,    both   in    semi--analytical   models
(\citealp{white&frenk91,     somerville&primack99,    kauffmannetal99,
coleetal00, hattonetal03,  monaco04, bertoneetal05}) and  in numerical
simulation          (\citealp{cen&ostriker91,         navarro&white93,
mihos&hernquist94,  katzetal96, thacker&couchman00, scannapiecoetal01,
springel&hernquist03,   rasera&teyssier06})   of   galaxy   formation.
Observational evidence for galactic outflows have already been pointed
out   by  several   authors   (\citealp{bland&tully88, heckmanetal90,  ellisonetal00,
heckmanetal00,    pettinietal01,    pettinietal02,   adelbergeretal03,
boucheetal06}).   They are usually  associated to  massive starbursts,
for which  very strong  outflows are reported:  for one solar  mass of
stars formed  in the  galaxy, between 1  and 5  solar mass of  gas are
ejected in the winds of starburst dwarf galaxies \citep{martin99}.

Despite the  difficulties of modelling supernovae  explosions within a
turbulent, multiphase and magnetized ISM, understanding the physics of
the resulting large  scale outflows is also a  challenge.  As explored
by  \cite{fujitaetal04}  in the  context  of  an isolated,  pre-formed
galactic  disc, the ram-pressure  of infalling  material might  be the
main limiting  factor for galactic winds to  exist.  The gravitational
potential  is  likely to  play  a minor  role  in  the development  of
galactic  winds,  especially in  low-mass  galaxies  where the  escape
velocity  is far  lower  than the  velocity  of the  wind. This  fact,
outlined first  by \cite{fujitaetal04}, is in  contradiction with most
semi--analytical studies  so far, for which only  the escalpe velocity
of the  parent halo determines if a  wind can develop or  not. In this
paper, using a self--consistent modelling of gas cooling and accretion
from the  hot surrounding  medium into the  disc, we  will demonstrate
that it is  indeed the ram pressure of the  infalling gas that matters
for galactic outflows.

To  illustrate  further   the  difficulties  in  modelling  supernovae
feedback, we  note that the  mass ejection rates measured  in galactic
winds obtained  in numerical simulations are  widely discrepant.  Some
authors  report  very  efficient  winds,  even in  the  quite  massive
galaxies (\citealp{springel&hernquist03, tasker&bryan06}), while other
authors  obtained  very  low  ejection  efficiency,  mainly  in  dwarf
galaxies \citep{maclow&ferrara99}.  We  believe that these differences
arise partly  from the different  numerical technics used,  and partly
from the different boundary  conditions imposed around the galaxy.  In
this  paper,  we  report  the  first  self--consistent  adaptive  mesh
refinement     (AMR)     simulation     of    a     collapsing     NFW
(\citealp{navarroetal96}) spherical cloud forming a galactic disk from
the  inside out.  In  the  recent literature,  this  same problem  was
adressed  using  smooth   particle  hydrodynamics  (SPH)  technics  by
\cite{springel&hernquist03}.    Grid--based   simulations  were   also
performed,   but   using    isolated,   pre--formed   rotating   discs
(\citealp{fujitaetal04, tasker&bryan06}).   One of the  key outcome of
such numerical  simulation is the fraction of  the injected supernovae
energy which is transferred up to the large scale galactic wind and is
truely     available      for     feedback     (\citealp{fujitaetal04,
scannapiecoetal06}).

We are indeed  facing a major difficulty when  introducing feedback in
numerical   simulations    that   include   radiative    losses:   the
characteristic time scale of  radiative cooling in dense regions where
star formation occurs is very fast, usually faster than the simulation
time step, so that one relies on implicit time integration schemes. As
a consequence,  if one handles  supernovae heating by  dumping thermal
energy in the  same dense regions, most (if not all)  of the energy is
radiated  away before any  large scale  flow develops.   This spurious
effect  has been  identified for  a  long time,  both in  cosmological
simulations       (\citealp{katz92},      \citealp{mihos&hernquist94},
\citealp{katzetal96})   and   in   interstellar   medium   simulations
(\citealp{wadaetal00},   \citealp{joung&maclow06}).   We   follow  the
recipe     proposed    by     \cite{morietal97}    and
\cite{gnedin98}  by  adding  around  each  supernova  event  density,
velocity and energy  profiles based on the Sedov  blast wave solution,
in order to get the correct  kinetic to thermal energy fraction in the
flow.  In  this way,  a large fraction  of kinetic energy  is injected
into the  disc, without  being radiated away  on the spot.   This will
translate  into a turbulent  gaseous disc,  and in  some cases  into a
large  scale  outflow.  To  account  for  the  small scale  effect  of
supernovae   feedback,   we   follow   the  multiphase   approach   of
\citealp{yepesetal97},      \citealp{ascasibaretal02},
\citealp{springel&hernquist03}, and \citealp{marri&white03}, for which
the supernovae thermal energy is  collecticely acccounted for as a new
equation  of state  for  the dense  ISM,  depending on  the gas  local
density (polytropic equation of state).  In this way, we can reproduce
a stiffer  equation of  state for the  dense gas, leading  to thicker,
more  stable  discs  (\citealp{toomre64},  \citealp{robertsonetal04}).
Another  problem  arises   because  the  scales  where
radiative  losses  of  SNe  remanants  occur  are  largely  unresolved
($\lesssim  1 \rm \,  pc$).  It  is therefore  likely that  the energy
transfer of SN  remnants to the ISM is  much lower (\citealp{larson74,
thorntonetal99})  than   the  energy  transfered   from  supperbubbles
initiated   by    multiple   localized   SNe   (\citealp{dekel&silk86,
melioli&gouveia04}).   In   reality,  modeling  properly   the  energy
deposition  requires  to  resolve  the  coherence length  of  the  SNe
explosions.  Thus, we have to bear in mind that our supperbubble model
entails many assumptions.

Within  this framework,  we  would  like to  answer  the  following
questions: What  is the conversion efficiency between the  smale scale
supernovae luminosity  and the  large scale outflow  luminosity~? What
are the conditions for a galactic  wind to develop and escape from the
parent halo potential well~?  What is the mass ejection rate of such a
wind~?   What is  the metallicity  of  the wind  and other  associated
observational signatures~?

Although the cooling  NFW halo approach that we  present in this paper
represents  an  important  improvement  over  previous  isolated  disc
grid-based  simulations,  we  are  still  a long  way  from  realistic
cosmological simulations. Such  idealized simulations are nevertheless
useful  to  understand the  physical  processes  responsible for  the
apparition of a galactic wind subject to boundary conditions which are
relevant for cosmology.  They also give us the possibility to derive a
simple analytical  model that highlights  the importance of  infall in
the process of wind formation.   Our paper is organized as follows. In
section \ref{anaprb} we present our analytical model and introduce the
relevant  quantities  for  constraining  the galactic  wind  formation
epoch.  In section \ref{nummeth},  we present our simulation settings,
with the physics  involved in our different disc  simulations, as well
as some  numerical details.  In section \ref{results},  we present our
main result,  namely the formation of  either a galactic  outflow or a
galactic fountain. We compare the actual wind formation epoch with our
analytical predicition. We also describe in details the flow structure
in each case, and discuss its observational signature.

\section{Galactic winds from cooling NFW halos}
\label{anaprb}

\begin{figure}
\centering{\resizebox*{!}{8.5cm}{\includegraphics{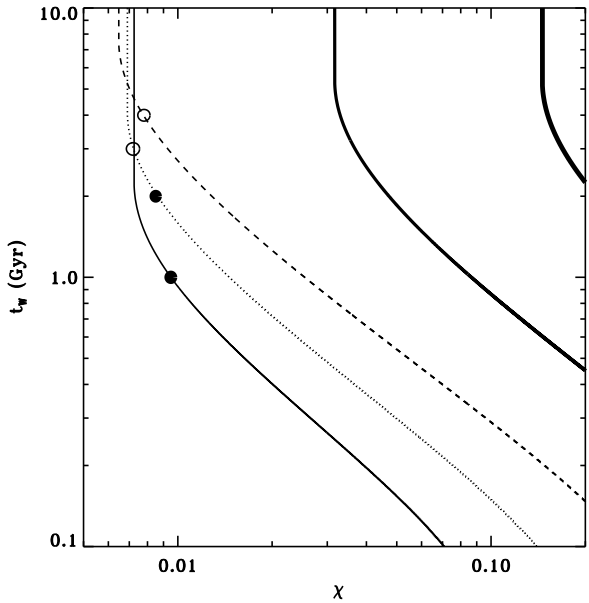}}}
\caption{The epoch of wind formation $t_W$ in Gyr as a function of the
hydrodynamical efficiency  $\chi$. We have assumed three different average star formation time $<t_*>=1 \,  \rm Gyr$ (thin solid line), $<t_*>=2 \,  \rm Gyr$ (dotted line) and $<t_*>=4 \,  \rm Gyr$ (dashed line) for the $10^{10}  \,  \rm  M_{\odot}$ halo, and the same $<t_*>=3 \,  \rm Gyr$ for the $10^{11}  \,  \rm  M_{\odot}$ (solid line) and the $10^{12} \, \rm M_{\odot}$ (thick solid line) halos. Open circles are for Sc ($t_W=3\, \rm Gyr$) and Sd ($t_W=4\, \rm Gyr$), circles are for Sa ($t_W=1\, \rm Gyr$) and Sb ($t_W=2\, \rm Gyr$).  }
\label{tvsetaw}
\end{figure}

In this section, we will describe the simplified problem we would like
to  study using high-resolution  numerical simulations,  as well  as a
simple  analytical  model  that  will  guide us  in  interpreting  the
results. We  model an  isolated NFW halo  (\cite{navarroetal96}), with
initially  similar  distribution  for  the  gas and  the  dark  matter
components (\citealp{ascasibaretal03}). We follow the initial settings
proposed by  \cite{springel&hernquist03}, so our work  can be directly
compared to them. Imposing indentical profiles for the gas and for the
dark matter  is especially  important in the  outer part, in  order to
have realistic accretion rates at late times, when the outer region of
the halo are  falling into the inner disc.   Both density profiles are
therefore given by
\begin{equation}
\rho={\rho_s \over r_s(1+r/r_s)^{2}} \, ,
\end{equation}
so that the total integrated mass is
\begin{equation}
M(<r)=4 \pi \rho_s r_s^3\left( \ln(1+r/r_s)-{r/r_s\over 1+r/r_s} 
\right) \, ,
\end{equation}
$\rho_s$  is the  characteristic density  of  the halo  and $r_s$  its
characteristic  radius.  We  define here  the Virial  radius as  the radius  where  the average
density is  equal to  200 times the  {\it critical}  density, assuming
$H_0=70$ km/s/Mpc.  We consider  a constant  gas fraction  of 15\%
throughout the  halo.  The temperature  profile is adjusted so  that a
strict hydrostatic equilibrium of the  gas sphere is maintained in the
non-cooling, non--rotating  case.  The halo  is truncated at  2 Virial
radii: outside the halo, we imposed a constant gas density equal to a
small  fraction ($10^{-4}$)  of  the density at this radius. Note that  this
truncation radius is a key parameter in this study. The size of the box is taken equal to 6 Virial radius. The concentration
parameter  $c=r_{vir}/r_s$ of  the NFW  profile is  taken to  be $10$,
independant of the halo mass.  Initially, our halo is slowly rotating,
with an angular momentum  profile corresponding to the average profile
found in 3D  cosmological simulations \citep{bullocketal01}, for which
the specific angular momentum is
\begin{equation}
j(r)=j_{max} {M(<r) \over M_{vir}}
\label{ang_mom}
\end{equation}
with     a     spin      parameter     $\lambda=J     \vert     E\vert
^{1/2}/(GM^{5/2}_{vir})$,   for    which   we   consider    only   two
representative  values  $\lambda=$0.04 and  0.1.   Starting from  this
initial equilibrium  configuration, the gas is allowed  to radiate its
thermal   energy  using   standard,  {\it   metal--dependant}  cooling
processes.  Note that initially we assume a constant metallicity equal
to  $10^{-3}$  solar  throughout  the  halo.  In  agreement  with  the
standard picture  of galaxy formation, a  centrifugally supported disc
quickly forms  in the halo  centre, gradually growing from  the inside
out.  Its size  depends strongly  on the  chosen spin  parameter. When
enough dense  material is  formed within the  disc, star  formation is
allowed to proceed  in a quiescent way, using  a star formation recipe
based   on  a  standard   Schmidt  law   (see  Eq.~\ref{schmidt_law}).
According  to \cite{salpeter55}  IMF,  for each  solar  mass of  stars
formed,  $\eta_{SN}  \simeq  0.1$   solar  masses  are  recycled  into
supernovae ejecta,  and drive strong shock waves  into the surrounding
ISM material.

Our  goal   is  to  implement  these  two   rather  standard  physical
ingredients in  the RAMSES  code, in order  to study the  formation of
galactic winds.  This will  be described in  the next section.  In the
present section,  we would like  to develop a simple  analytical model
that allows us  to capture the basic mecanism  driving galactic winds.
Our picture  contrasts with the one  discussed in \cite{fujitaetal04},
who   assumed   a   pre-formed   gaseous  galactic   disc,   with   no
self-consistent star formation. A blast wave was initiated at the very
center of the disc, with an energy budget consistent with massive star
bursts. Depending on the blast  wave and disc parameters, two possible
scenarii  emerge:  ``blow  out''  for  which the  disc  is  completely
destroyed,  and  ``blow  away'',  corresponding  to  a  galactic  wind
escaping the  galactic disc. These authors have  demonstrated that the
main limiting factor for a wind to appear in the ``blow away'' case is
the amount of infalling gas,  coming from the surrounding halo. In the
present  study,  we  will  address  this question  using  a  different
approach, based on quiescent star  formation and infall from a cooling
NFW halo.

Let  us consider  that the  gas  is cooling  instantaneously, so  that
pressure  support is  lost from  the very  beginning.   Each spherical
shell is  therefore free-falling , gradually  accelerating towards the
center. We compute  the accretion rate of halo  material onto the disc
using         a        standard         semi--analytical        recipe
(e.g. \citealp{hattonetal03}), namely by taking the time derivative of
the integrated accreted mass
\begin{equation}
M_{acc}(<t) = M \left(<r_{ff}(t)\right) \, ,
\end{equation}
where  $r_{ff}(t)$  is  the  radius  for  which  the  free--fall  time
$t_{ff}(r)$ is  equal to the current  time $t$. The  gaseous disc mass
therefore  grows by accretion  and decays  because of  star formation,
leading to a simple differential equation for its time evolution
\begin{equation}
\dot M_g = \dot M_{acc}-{M_g \over <t_*>} \, .
\end{equation}
Here,  $<t_*>$ is  the  (mass--weighted) average  star formation  rate
within  the disc.  It  depends on  the exact  disc geometry  (size and
thickness)  and  on the  Schmidt  law  parameters.  As a  first  order
approximation, one can show that 
\begin{equation}
 <t_*> \propto t_0 \lambda_0 \, :
\label{tstar}
\end{equation}
small  discs form  stars  more  efficiently than  large  ones. If  one
assumes that $<t_*>$ remains constant  in time, one gets the following
formal solution
\begin{equation}
M_g(t) = \int _0 ^t \dot M_{acc}(u)\exp (\frac{u-t}{<t_*>}) 
\mbox{d}u\, .
\label{mgas}
\end{equation}
We finally deduce the total supernova luminosity in the disc using
\begin{equation}
L_{SN}=\eta_{SN}\epsilon_{SN}\dot M_* = 
\eta_{SN}\epsilon_{SN}{M_g \over <t>_*} \, ,
\end{equation}
where $\epsilon_{SN} \simeq 50$ keV is the specific energy produced by
one single supernova ($10^{51}$ erg  for an average progenitor mass of
10 $M_{\odot}$). Part of the supernova energy contributes to sustain a
turbulent multiphase  ISM. Only a  (small) fraction of this  energy is
converted into actual wind  luminosity escaping from the dense gaseous
disc.  The  fraction of  energy that manage  to escape from  the dense
gaseous disc will depend  on the disc characteristic (thickness, size,
gas content).  In our  simulations, these properties will be specified
by the the  spin parameter of the halo and by  the star formation time
scale  of the  Schmidt  law. We  parametrize  this unknown  conversion
factor $\chi$ by defining the  wind luminosity as $L_W = \chi L_{SN}$.
Our goal is to perform  numerical simulations of star forming discs in
order  to  compute the  actual  conversion  efficiency resulting  from
various, complex  hydrodynamical processes  within the disc.   We thus
call this parameter $\chi$ the {\it hydrodynamical efficiency}.

Our goal is  now to derive under which conditions  a galactic wind can
form   within    the   halo.    We   know    from   previous   studies
(\citealp{fujitaetal04}) that, in presence  of infalling gas, the main
factor preventing the wind from blowing out is the ram pressure of the
surrounding material.   This is a much more  stringent constraint than
any criterion based  on the escape  velocity.   If we assume that  the free-falling
cold gas  hits the disc plane  with terminal velocity,  we can compute
the accretion luminosity as
\begin{equation}
L_{acc}(t)={1\over2} V_{0}^2(r_{ff}(t))\dot M_{acc}
~~~\mbox{and}~~~
V_{0}^2(r)=-2 \int_{0}^{r} {G M(x) \over x^2} dx
\end{equation}
We define  the {\it wind beak-out  epoch} $t_W$ as the  time for which
the    wind    luminosity    exceeds    the    accretion    luminosity
$L_{acc}(t_W)=L_W(t_W)$.  In  Figure~\ref{tvsetaw}, the wind break-out
epoch is shown  for various halo masses, as a  function of our unknown
parameter   $\chi$.    One  clearly   sees   that   the  smaller   the
hydrodynamical efficiency,  the later  the wind will  blow out  of the
disc.  More importantely, {\it for  each halo mass, there is a minimum
efficiency below which  no wind can form at  all}.  This minimal value
for  $\chi$ is  around 0.7\%  (resp.  3\%  and 15\%)  for  a $10^{10}$
M$_\odot$   halo    (resp.    $10^{11}$   M$_\odot$    and   $10^{12}$
M$_\odot$). We  conclude from  this very simple  toy model that  it is
critical  to   determine  the  actual   hydrodynamical  efficiency  in
quiescent star  forming galaxies.   Our goal is  here to  perform high
resolution numerical simulations of  isolated galaxies, to measure the
epoch  when   galactic  winds  appear   and  deduce  from   $t_W$  the
corresponding (most likely) hydrodynamical efficiency.

\section{Numerical methods}
\label{nummeth}

Our simulations were performed with the Adaptive Mesh Refinement (AMR)
code RAMSES \citep{teyssier02}. The  gas evolution is computed using a
second--order   Godunov  scheme   for  the   Euler   equations,  while
collisionless  star  particles'  trajectories  are  computed  using  a
Particle--Mesh solver. The dark matter component is accounted for as a
constant background gravitational potential. Gas cooling is taken into
account as a source term  in the energy equation. The cooling function
is computing  using the \cite{sutherland&dopita93} cooling model,  using a
look-up  table  in  the  temperature and  metallicity  plane.   Metals
created in the surrounding of supernova explosions are advected by the
hydrodynamics solver as a passive scalar.

We now describe  in more details our 3  main physical ingredients used
for  the  present  study,  namely star  formation,  supernova  thermal
feedback and supernova kinetic feedback.

\subsection{Star formation recipe}

In each cell, gas is converted into star particles following a Schmidt
law
\begin{equation}
\dot \rho_* = - {\rho \over t_*} \; {\rm if} 
\; \rho>\rho_0 \, , ~~~~\dot \rho_* = 0 \; {\rm otherwise} \, ,
\label{schmidt_law}
\end{equation}
where  $\rho_0$ is  an arbitrary  density threshold  defining  what we
consider here as being ``interstellar medium''. The star formation
time scale is proportional to the local free-fall time, 
\begin{equation}
t_*=t_0 \left ( {\rho \over \rho_0} \right ) ^{-1/2} \, .
\end{equation}
The   two   parameters   $\rho_0$   and   $t_0$   are   poorly   known
and scale  dependent. One standard  approach consists
in calibrating  these numbers to  the observed star formation  rate in
local   galaxies,  the   so--called   \cite{kennicutt98}  law,   which
translates roughly into $\rho_0 \simeq 0.1 \, \rm H . \rm cm^{-3}$ and
$t_0 \simeq 1-10 \, \rm Gyr$.  In the present study, we allow $t_0$ to
take    2    values,    namely    $3$    or    $8    \,    \rm    Gyr$
corresponding   to  star  formation   efficiencies  of
respectively  $\sim 0.05$ and  $0.02$, last  value is  compatible with
\cite{krumholz&tan07}  if we  suppose that  star  formation efficiency
remains the same  at very high ($>  10^5 \, \rm H .  \rm cm^{-3}$) and
unresolved densities  (maximum density in our simulations  stalls to a
few $ 10^1 \, \rm H . \rm  cm^{-3}$ and $ 10^2 \, \rm H . \rm cm^{-3}$
respectively for  the $10^{10}$ M$_\odot$ and  $10^{11}$ M$_\odot$ due
to polytropic  equation of state).  When  a gas cell  is eligible for
star  formation, N  collisionless  star particles  are
spawned using a Poisson random process, with probability
\begin{equation}
P(N)={\lambda_P \over N !} \exp({-\lambda_P}) \, ,
\label{poisson_law}
\end{equation}
where the mean value is 
\begin{equation}
\lambda_P= \left ( {\rho \Delta x^3 \over m_*}\right ) {\Delta t \over t_*} \, .
\label{poisson_param}
\end{equation}
The mass  of a  star particle is  taken to  be an integer  multiple of
$m_*=\rho_0      \Delta     x_{min}^3/(1+\eta_{SN}+\eta_{W})$     (see
Appendix~\ref{sn_scheme} for details  on $\eta_{W}$), and is therefore
directly     connected    to     the    code     spatial    resolution
\citep{rasera&teyssier06}. The minimum mass  of a single star particle
is therefore $\sim 500$ M$_\odot$,  but high density regions can spawn
more massive star  particles. We also ensure that  no more than $90\%$
of the gas in a cell is depleted by the star formation process.

\subsection{Polytropic equation of state}

\begin{figure*}
\centering{\resizebox*{!}{8.5cm}{\includegraphics{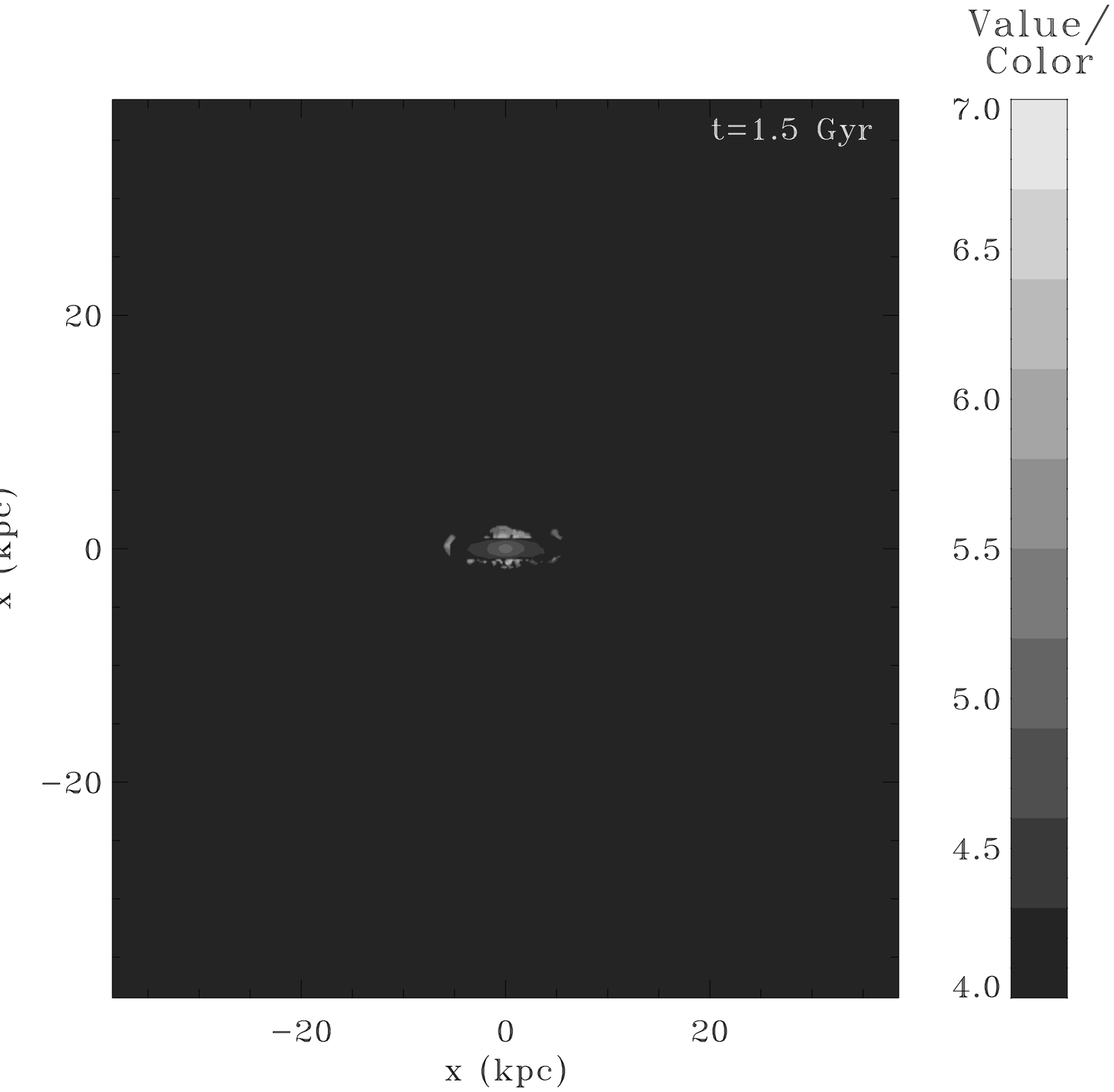}}}
\centering{\resizebox*{!}{8.5cm}{\includegraphics{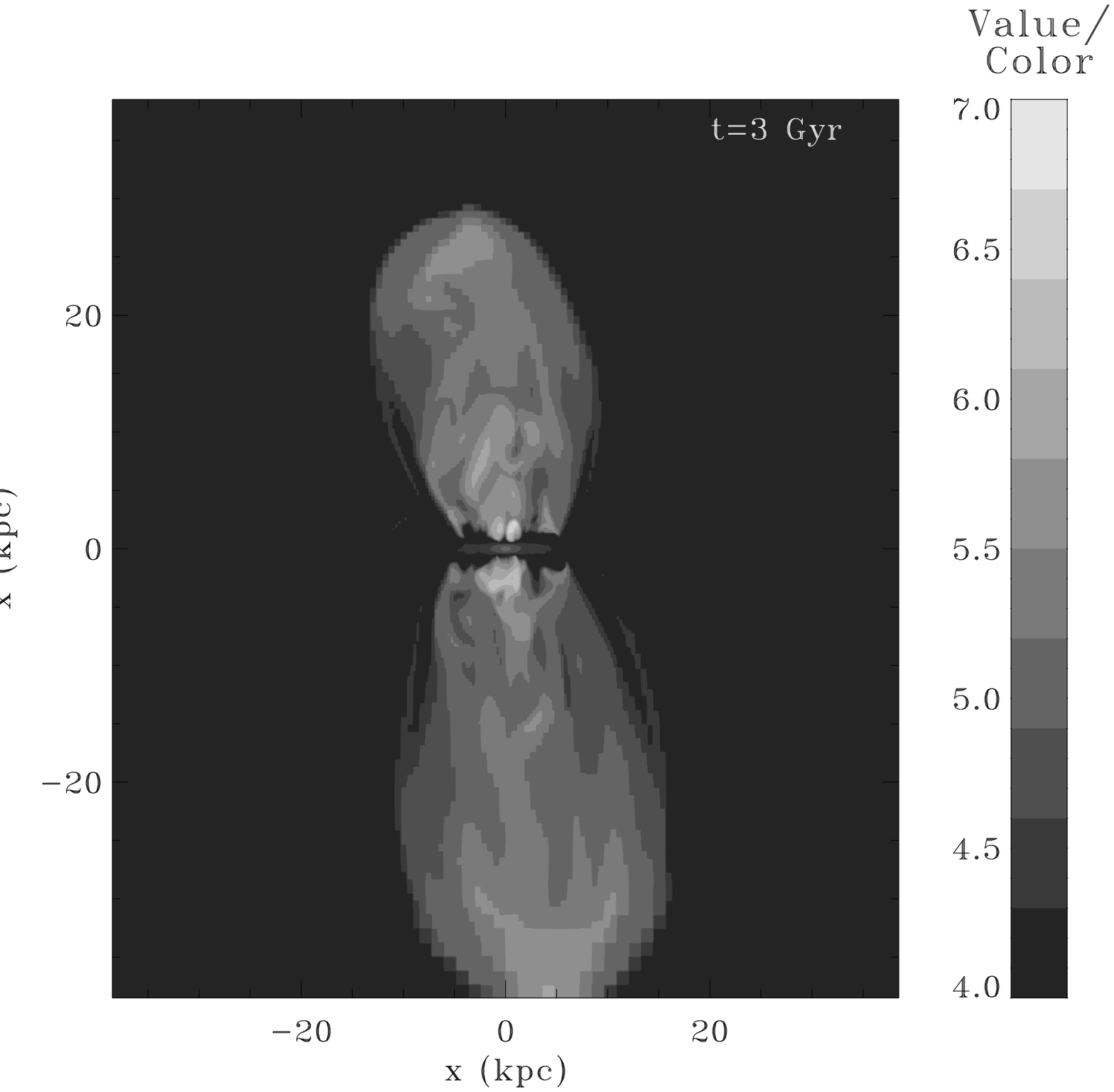}}}
\centering{\resizebox*{!}{8.5cm}{\includegraphics{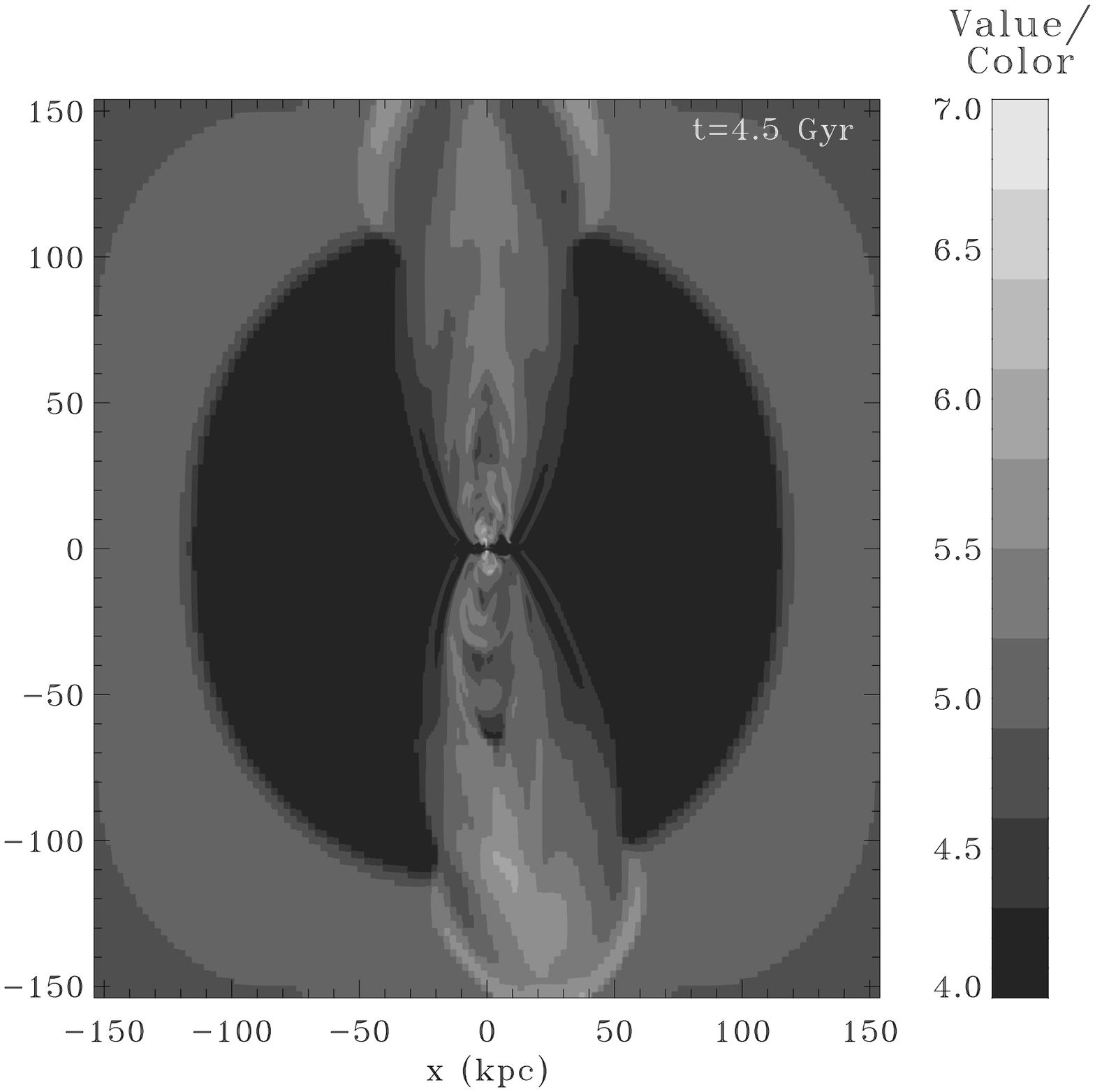}}}
\centering{\resizebox*{!}{8.5cm}{\includegraphics{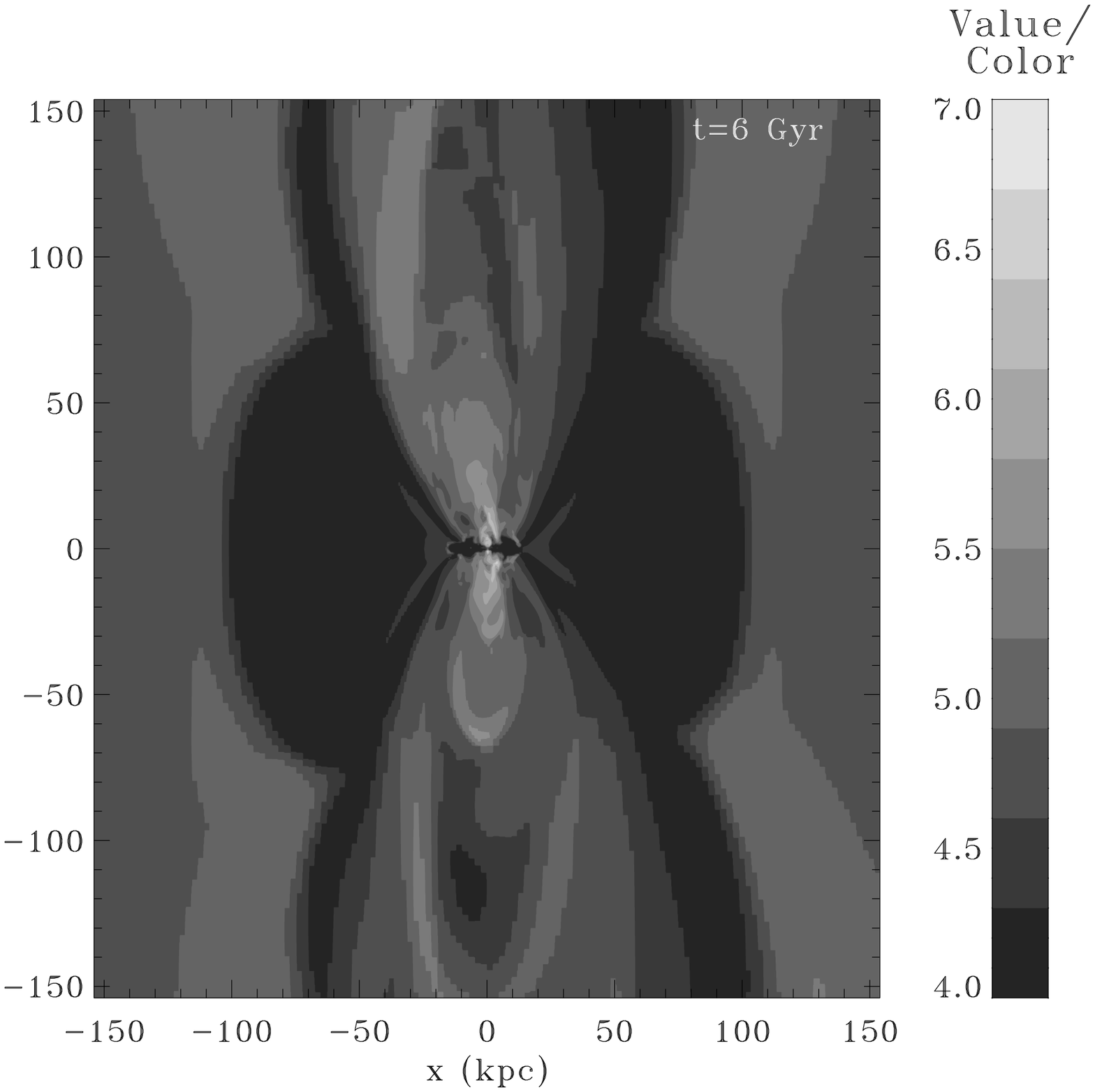}}}
\caption{Cut  of  the  gas  temperature  in  the  Oyz  plane  for  the
Sd simulation at different epochs.  The two top
pannels are  a zoom 4 times of  the simulation box and  the two bottom
pannels  are the  entire  simulation  box. The  colour  scale gives  the
temperature in $\log (\rm{K})$. Note that length scales are not the same in each pannel.}
\label{wind1d10_T}
\end{figure*}

The second  important ingredient for our galactic  disc simulations is
the  thermal  feedback  of  supernovae  into the  ISM,  based  on  the
multiphase       model       of       \cite{ascasibaretal02}       and
\cite{springel&hernquist03}.  The  idea is to  assume that the  ISM is
driven  by  small  scale  effects  (turbulence,  thermal  instability,
thermal  conduction, molecular cloud  formation and  evaporation) that
quickly reach  steady--state and lead to  a quasi--equilibrium thermal
state, for  which the  average temperature is  a function of  the mass
density  alone. In  practice, when  the  gas density  exceeds the  ISM
threshold $\rho_0$,  the gas  temperature is forced  to be equal  to or
higher than
\begin{equation}
T = T_0 \left ( {\rho \over \rho_0} \right ) ^{\gamma_0 -1} \,
\end{equation}
where $\gamma  _0$ is the polytropic  index, and can be  chosen to the
exact value  of the multiphase  model (\citealp{springel&hernquist03})
or to  a constant  value for  sake of simplicity.   If $\gamma  _0$ is
chosen between 4/3 et 2, as prescribed by \cite{springel&hernquist03},
the resulting  equation of  state gives similar  results for  the disc
thickness  and   maximum  density.    Our  standard  choice   is  here
$\gamma_0=5/3$.

\begin{figure*}
\centering{\resizebox*{!}{8.5cm}{\includegraphics{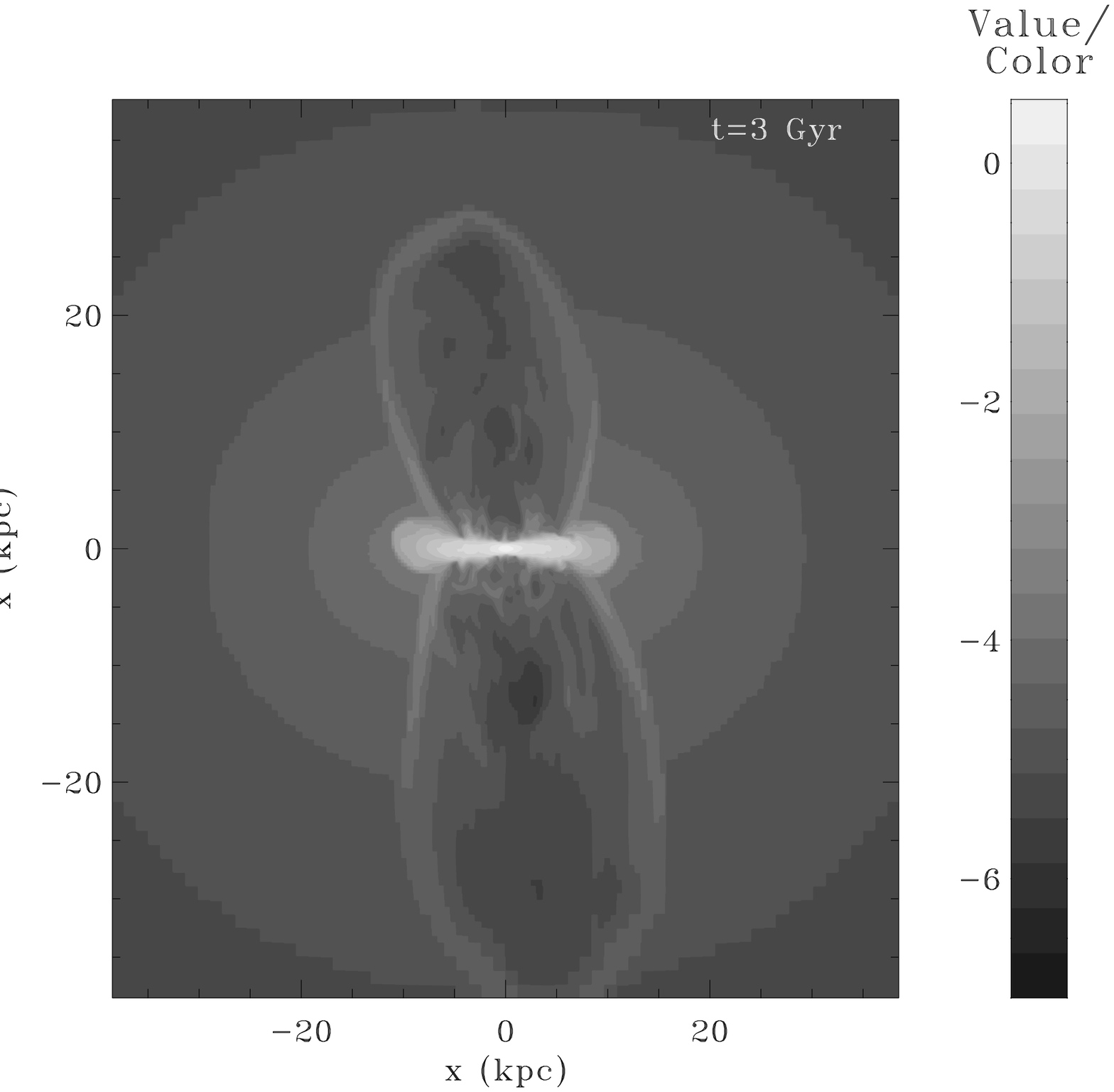}}}
\centering{\resizebox*{!}{8.5cm}{\includegraphics{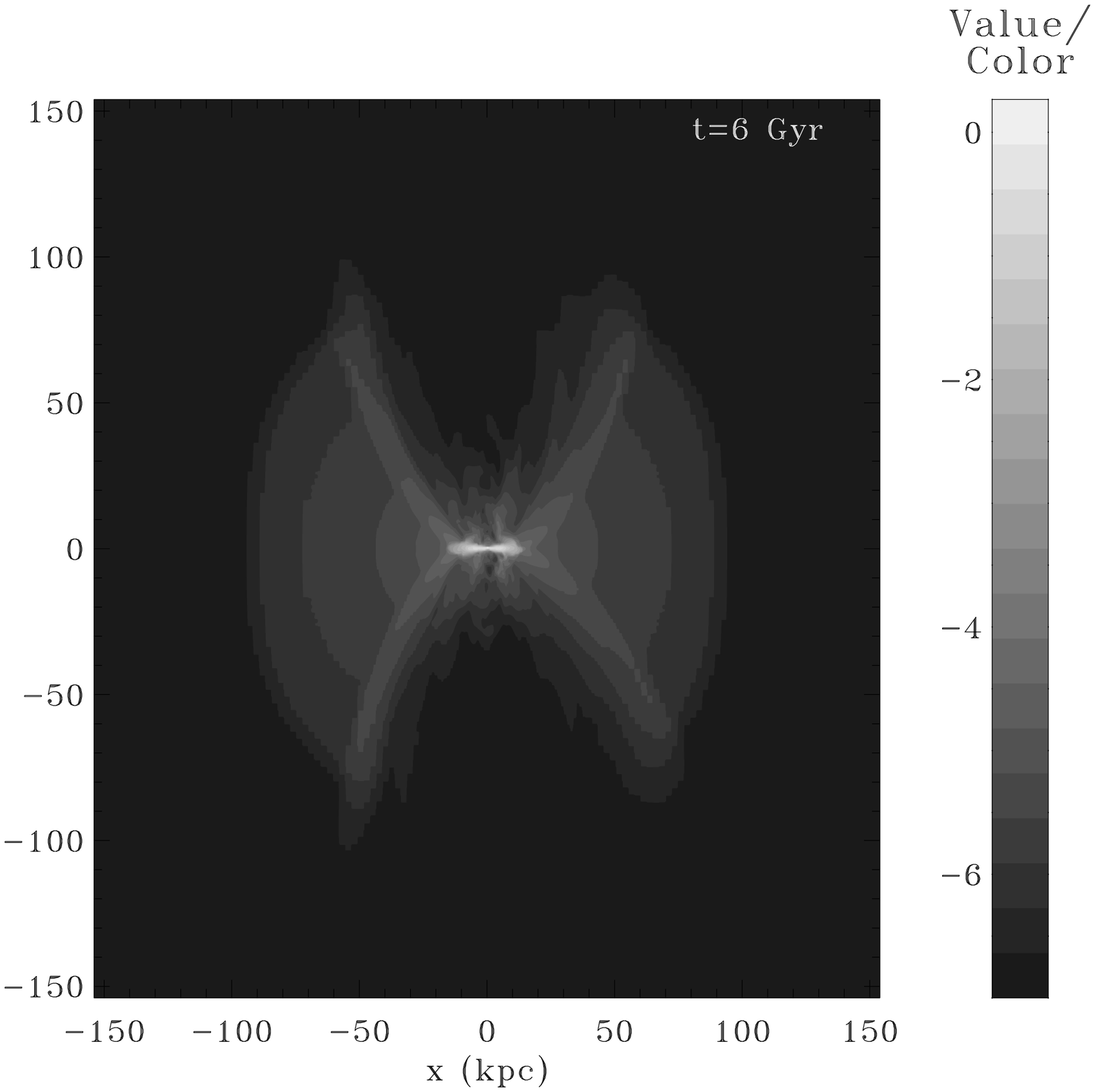}}}
\centering{\resizebox*{!}{8.5cm}{\includegraphics{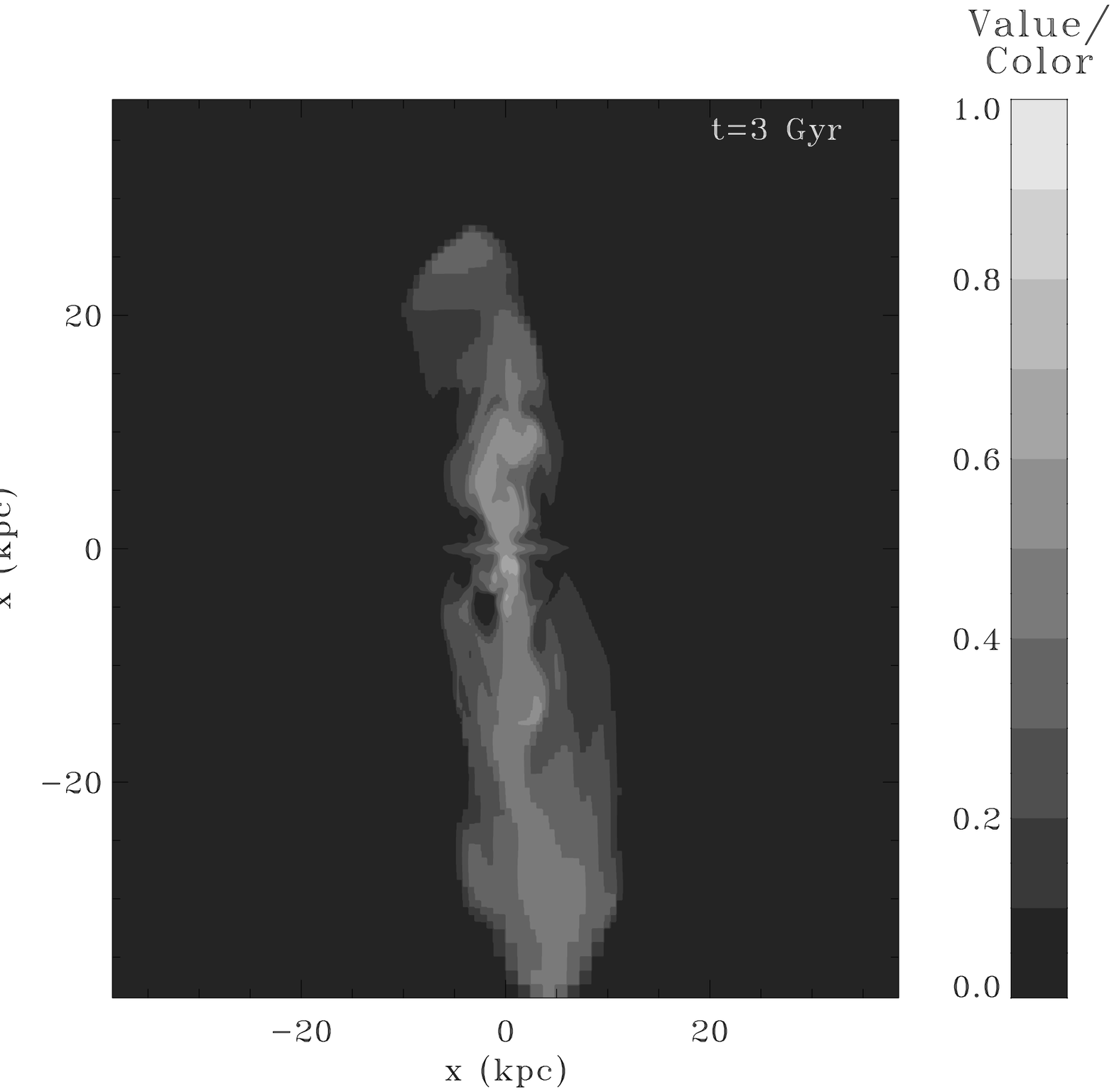}}}
\centering{\resizebox*{!}{8.5cm}{\includegraphics{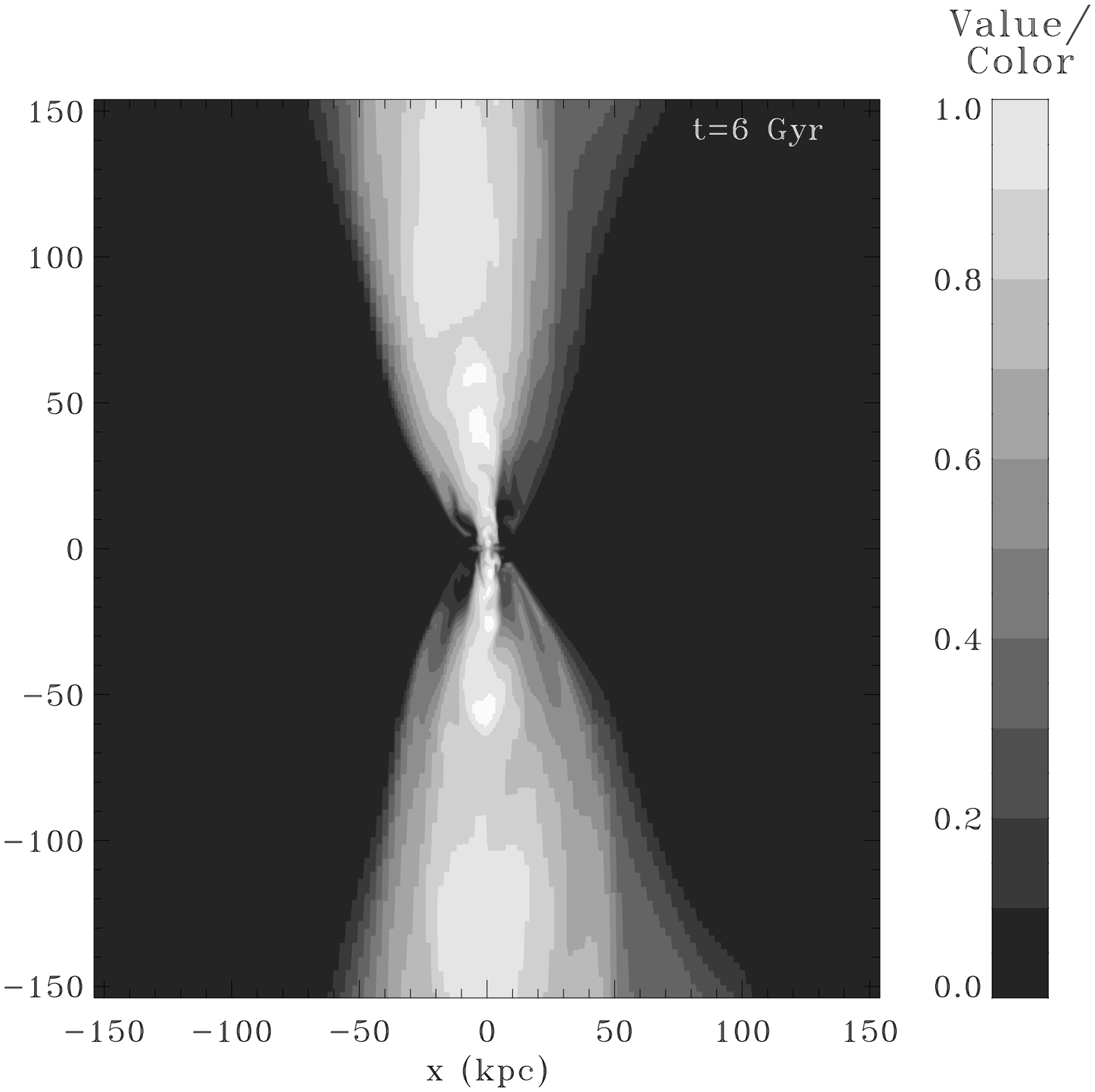}}}
\caption{Cut  of  the  gas density  (up)  and metallicity (bottom) in  the  Oyz plane  for  the Sd 
simulation at different epochs. The two left
pannels are  a zoom 4  times of the  simulation box and the  two right
pannels  are the  entire  simulation  box. The  colour  scale gives  the
density  in $\log(\rm  {cm^{-3}})$ (up)  and the  metallicity  in $\rm
Z_{\odot}$ (bottom). Note that length scales are not the same in each pannel.}
\label{wind1d10_dZ}
\end{figure*}

\subsection{Blast wave model}

The  main problem  of the  previous thermodynamical  approach  is that
feedback  is only  accounted for  at mesoscopic  scales. On  the other
hand,   it   is   now   an   observational  fact   that   very   large
supernovae--driven  bubbles  appear quite  naturally  in star  forming
galaxies.  These superbubbles are built by the coalescence of numerous
supernovae  remnants,  who   collectively  create  these  large--scale
features, driving  very large, macroscopic flows  that eventually give
rise to  galactic winds  or galactic fountains.   Current cosmological
simulations do not have the necessary resolution to resolve individual
supernovae blast wave. On the  other hand, they do have the resolution
to resolve superbubbles. Our goal  is here to inject kinetic energy in
the  form of  spherical blast  waves  of size  comparable to  galactic
superbubbles, namely $r_{SN} = 100$  to $200$ pc.  This kinetic energy
injection can be considered as a turbulent forcing term with injection
scale $r_{SN}$. 

It is now well-established that  supernovae energy can not be released
only by direct thermal energy injection ~\citep{navarro&white93}.  For
high  density star  forming region,  the  gas radiates  away all  this
thermal energy in  one time step due to very  fast atomic cooling. The
consequence  is that  SNe  have  no effect  on  the dynamics.  Various
methods have  been proposed to correctly incorporate  the SNII kinetic
feedback  into numerical  simulations. The  first idea,  discussed for
example  in \cite{governatoetal06} is  to artificially  stop radiative
cooling in the region where the SNII explosion occurs, for a time long
enough for the blast wave to develop and expand. This trick allows for
a large enough fraction of the SNII energy to be converted into actual
gas kinetic energy. The second  approach is to directly inject kinetic
energy into the  surrounding gas. \cite{springel&hernquist03} used for
that purpose high-velocity collisionless  particles that travel over a
rather large distance (up to a few kpc) before being re-incoporated in
the  fluid as  new SPH  particles.

In  this  paper, we  direclty  add  to  each flow  variable  (density,
momentum and total  energy) a spherical blast wave  solution of radius
$r_{SN} = 150$  pc, assuming that each supernovae releases half of its energy in this kinetic form (the other half is accounted for the polytropic equation of state). As soon as this distance  is large enough compared
to the grid resolution, we  minimize spurious energy losses. Each time
a ``star'' particle is created, we simultaneously remove an equivalent
gas  mass  from  the  cell   that  is  entrained  by  the  SNII  blast
wave. Recalling that $m_*$ is  the mass locked into long--lived stars,
the   total   mass   removed   from   each   star--forming   cell   is
$m_*(1+\eta_{SN}+\eta_W)$,  where the entrainement  parameter $\eta_W$
is defined  as in  \cite{springel&hernquist03}.  We have  used $\eta_W
\simeq 1$ in order to reproduce roughly a Sedov blast wave solution in
typical galactic  discs for which  the density is slightly  greater to
$\rho_0$. Note  that in order  to implement this numerical  scheme, we
have  used  ``debris'' particles  to  describe  the  Sedov blast  wave
profile  around  each  SNII  explosion.   Details  are  given  in  the
Appendix.  The most important parameter is $r_{SN}$, the radius of the
spherical blast wave.  This sets  the injection scale of the turbulent
cascade   in  the   disc   (\citealp{joung&maclow06}).   This   value,
comparable   to   the    typical   size   of   galactic   superbubbles
(\citealp{mackee&ostriker77}),  is   however  larger  than  individual
supernovae remmants, whose size just before the snow plow phase varies
from 10  to 50 pc, depending  on the exact early  shock wave dynamics,
whether  adiabatic or  evaporative (\citealp{cioffietal88}).   We have
not  tried to  study the  impact of  this important  parameter  on our
results,  in order to  avoid clouding  the purpose  of this  paper. We
rather keep this parameter fixed to its fiducial value of $r_{SN}=150$
pc (\citealp{mackee&ostriker77}, \citealp{shull&silk79}).

\subsection{Simulation parameters}

\begin{table}
\begin{center}
\begin{tabular}{|r|c|c|c|c|c|c|}
\hline
Run & $V_{vir}$ & $\lambda$ & $t_0$ & $\ell_{min}$ & $\ell_{max}$ & $\Delta x$\\
 & km/s &  & Gyr & &  & pc\\
\hline
Sa & $35 $ & 0.04 & 3 & 7 & 11 & 73 \\
\hline
Sb & $35 $ & 0.04 & 8 & 7 & 11 & 73 \\
\hline
Sc & $35 $ & 0.1 & 3 & 7 & 11 & 73 \\
\hline
Sd & $35 $ & 0.1 & 8 & 7 & 11 & 73 \\
\hline
La & $75 $ & 0.04 & 3 & 7 & 12 & 78 \\
\hline
Lb & $75 $ & 0.04 & 8 & 7 & 12 & 78 \\
\hline
Lc & $75 $ & 0.1 & 3 & 7 & 12 & 78 \\
\hline
Ld & $75 $ & 0.1 & 8 & 7 & 12 & 78 \\
\hline
\end{tabular}
\end{center}
\caption{Parameters used for the simulations performed in this paper.}
\label{tab1}
\end{table}

We use  for  the fluid  solver  simple ``outflow''  boundary
conditions,  for which  all flow  variable  are assumed  to have  zero
gradient at the boundary.  All  simulations are initialized with a NFW
gas and  dark matter  halo in hydrostatic  equilibrium, with  the same
density profile  for both  fluids.  The dark  matter component  is not
explicitly  simulated   here;  only   baryons  (stars  and   gas)  are
self--gravitating. We  therefore add a  fixed analytical gravitational
potential  to the  solution  of the  Poisson  equation.

We  have  performed our  simulations  for  2  different halo  circular
velocity,  namely $V_{vir}=35$  and  $75$ km/s,  which correspond  to
Virial  masses of  $10^{10}  \,  \rm M_{\odot}$  and  $10^{11} \,  \rm
M_{\odot}$  at redshift  zero.  We  have explored  two  different star
formation efficiency, with timescales $t_0=3 \, \rm Gyr$ and $t_0=8 \,
\rm Gyr$.  We use a coarse grid of $128^3$ cells, in order to properly
describe the large scale  dynamics within the halo, which corresponds,
in  the   RAMSES  terminology,  to  a  miminum   level  of  refinement
$\ell_{min}=7$. The grid  is further refined up to  4 or 5 additionnal
levels   of  refinement   ($\ell_{max}=11$  or   $12$ corresponding to $\sim 75 \, \rm pc$, see table~\ref{tab1} for more details),  based   on  a
quasi-Lagrangian strategy,  for which  a cell is  refined if  its mass
exceeds  a threshold of  $2\times 10^{-6}$  the Virial  mass. This refinement strategy verifies the Jeans length criterion inside the galactic disc because of the polytropic component that helps to stabilize against fragmentation.  Starting
with $2\times  10^6$ cells, our  simulations usually reach a  total of
$4$ to  $5\times 10^6$ cells.  We  end our simulation  after a typical
integration time of 6 Gyr,  which corresponds to the free-fall time of
the halo outer boundary at 2 Virial radius.

\begin{figure}
\centering{\resizebox*{!}{8.5cm}{\includegraphics{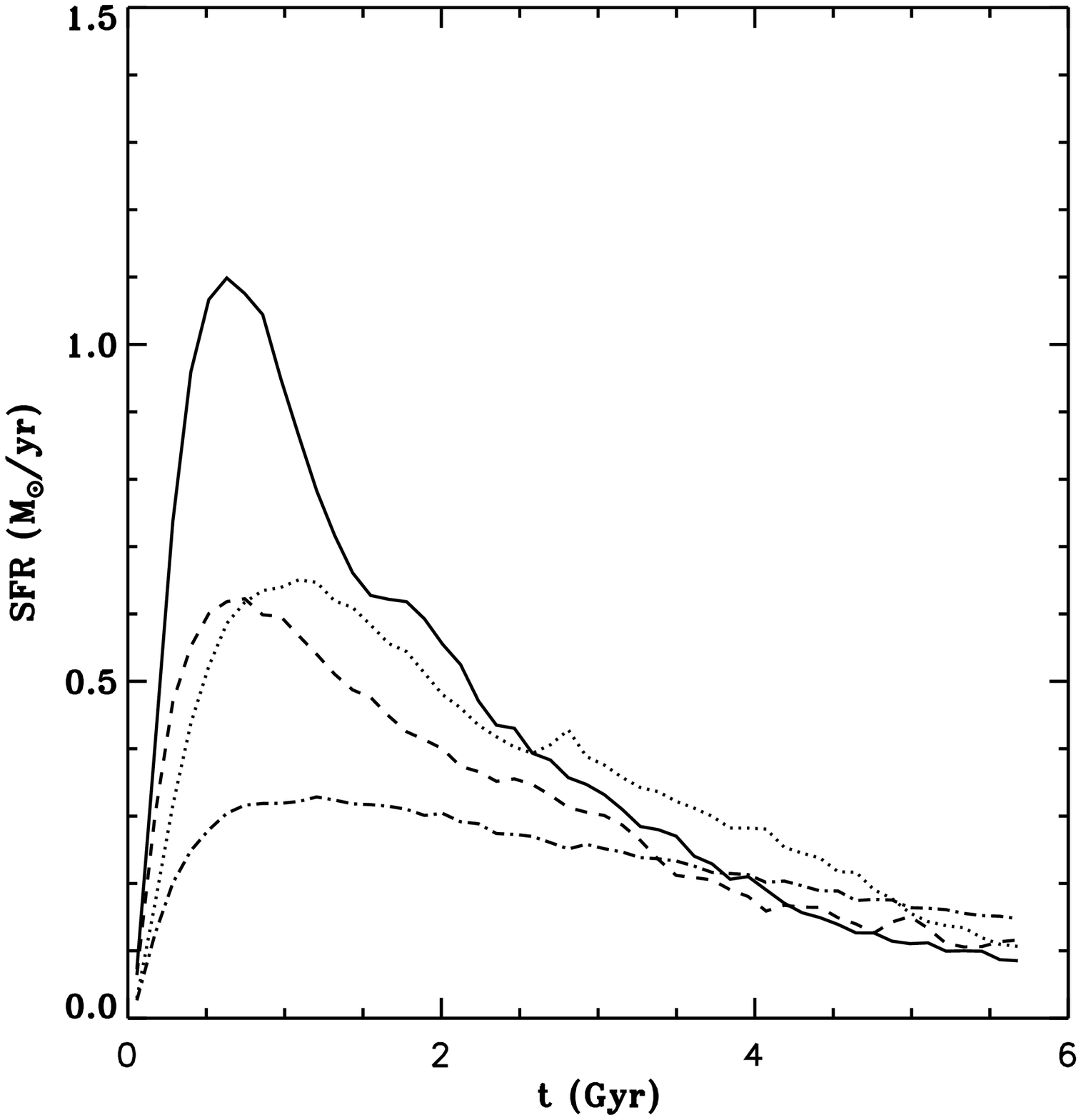}}}
\caption{Star formation rates of  the $10^{10} \, \rm M_{\odot}$ halos
with $\lambda=0.04$ and $t_0=3\, \rm Gyr$ (solid line), $\lambda=0.04$
and $t_0=8\,  \rm Gyr$ (dotted  line), $\lambda=0.1$ and  $t_0=3\, \rm
Gyr$   (dashed  line)   and  $\lambda=0.1$   and  $t_0=8\,   \rm  Gyr$
(dash-dotted line).}
\label{sfr1d10}
\end{figure}

\begin{figure*}
\centering{\resizebox*{!}{8cm}{\includegraphics{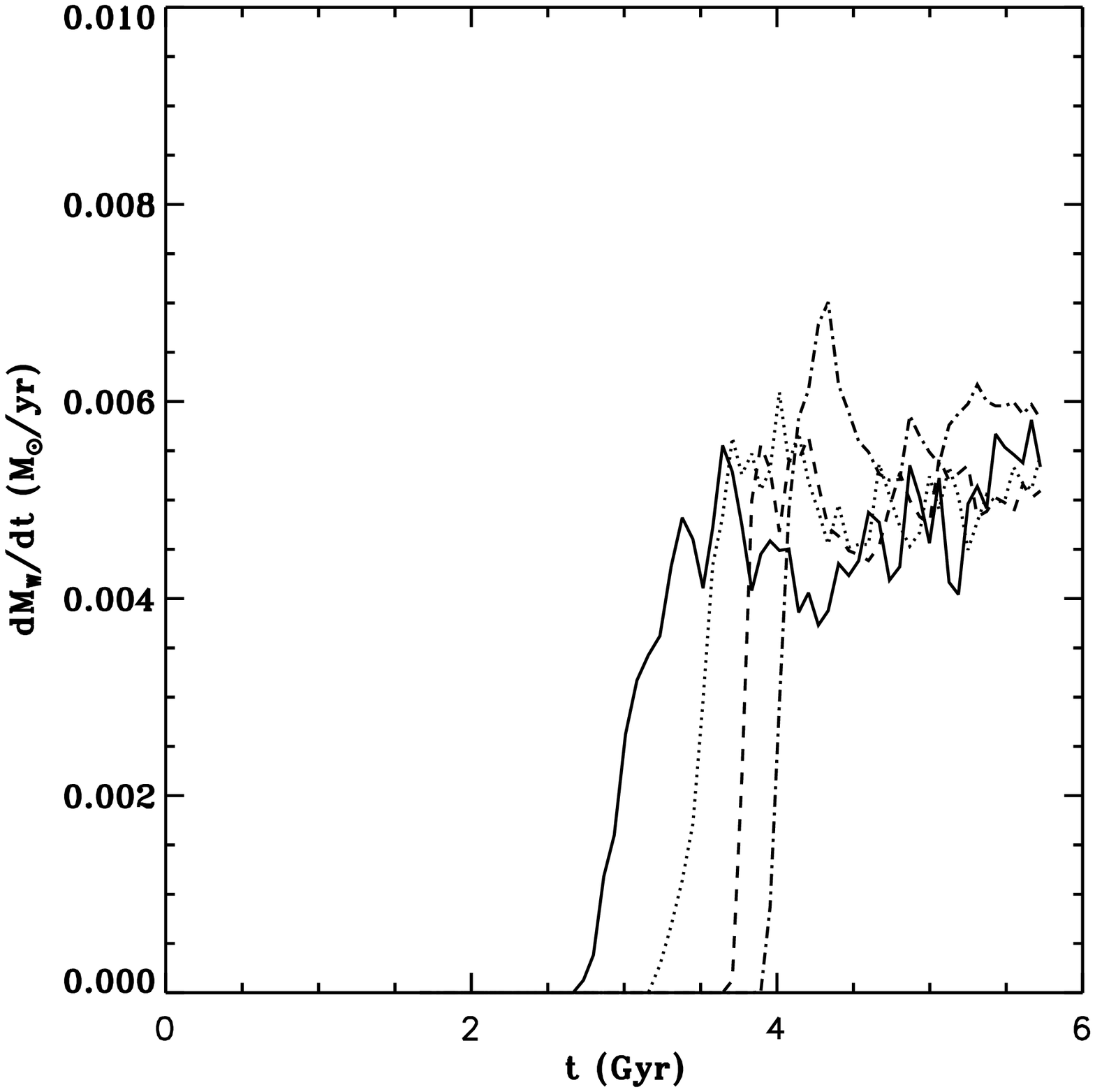}}}
\centering{\resizebox*{!}{8cm}{\includegraphics{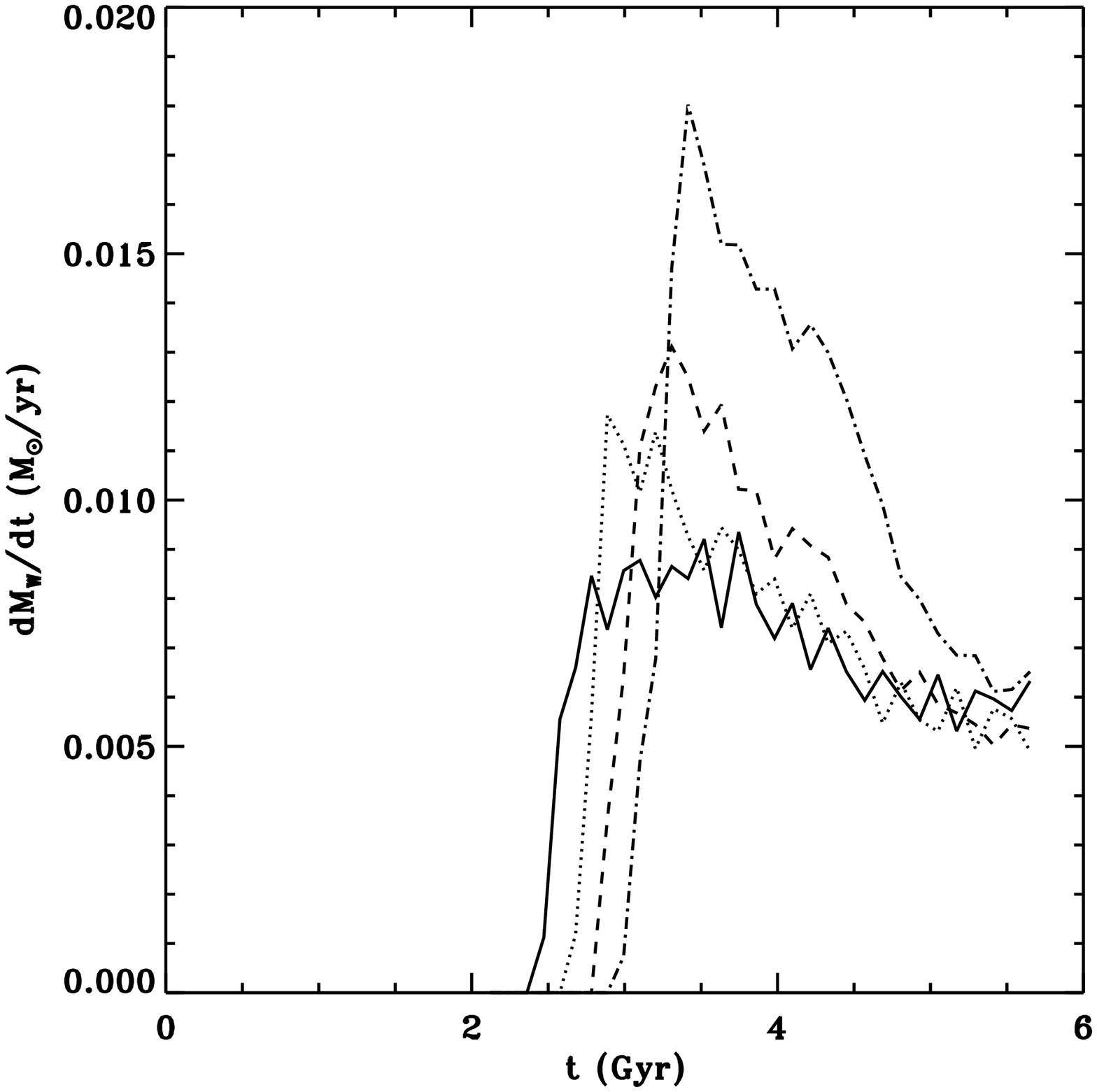}}}
\centering{\resizebox*{!}{8cm}{\includegraphics{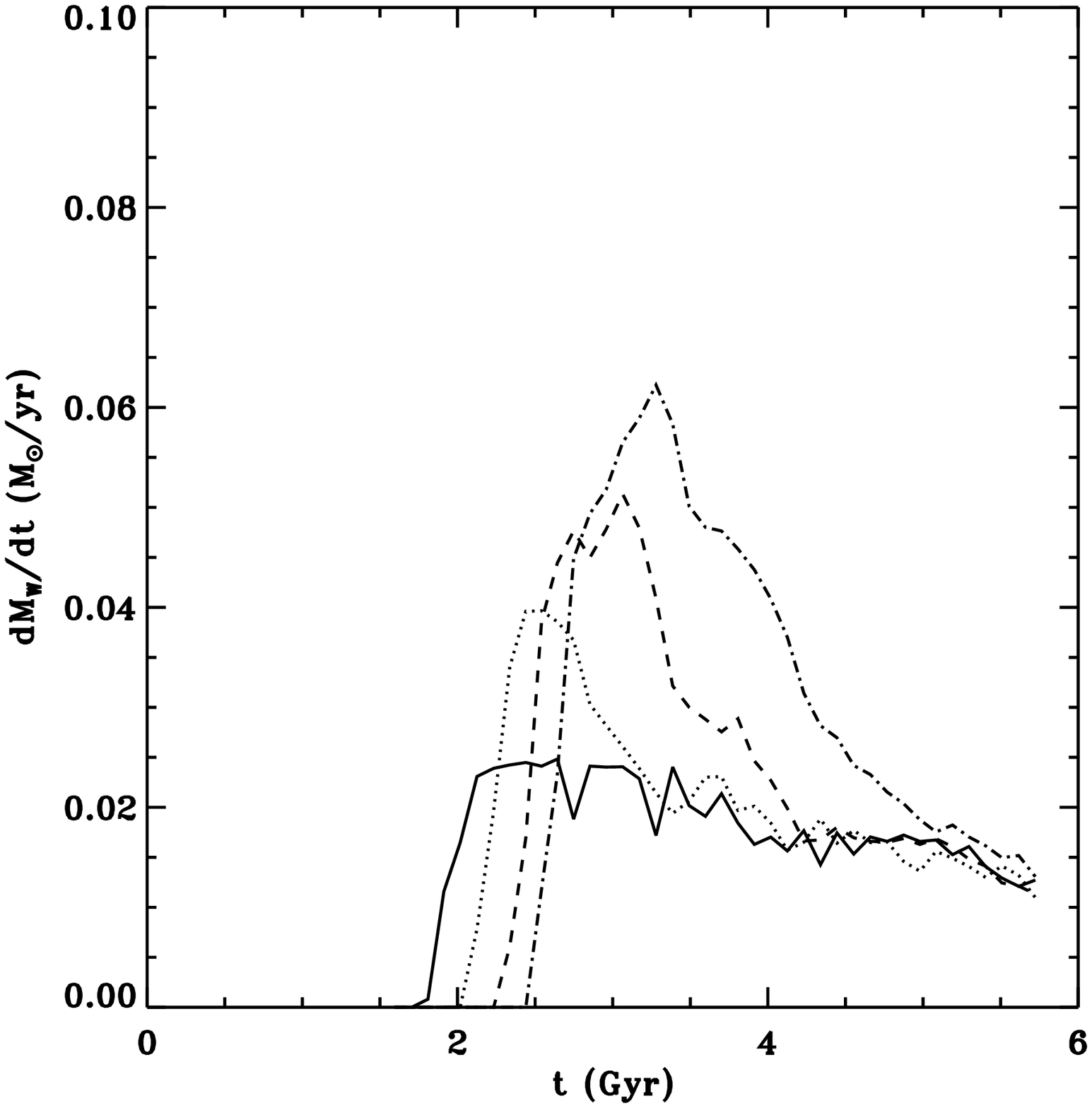}}}
\centering{\resizebox*{!}{8cm}{\includegraphics{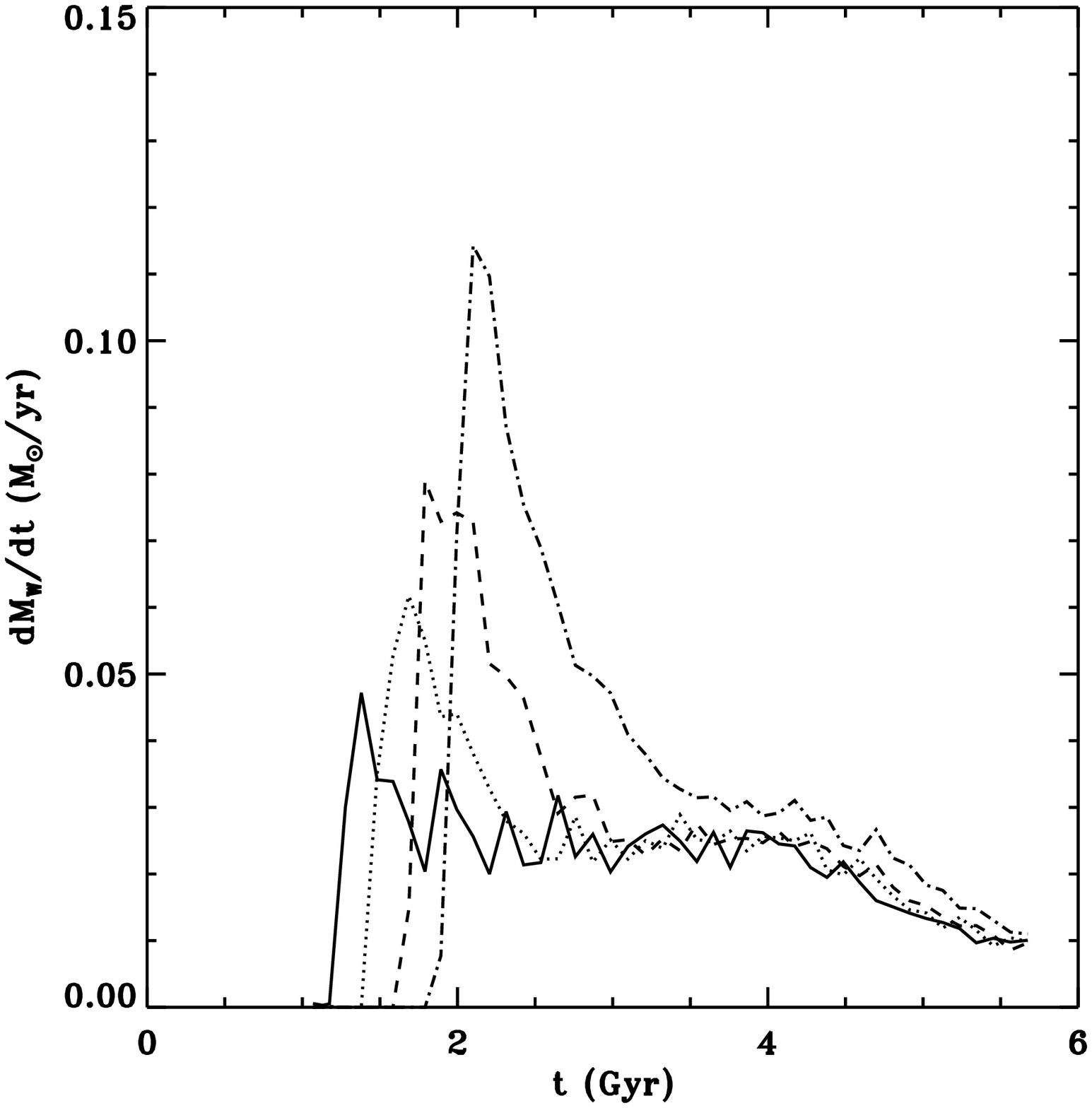}}}
\caption{Mean flux  of mass outflowing calculated for  the $10^{10} \,
\rm  M_{\odot}$ halos with  $\lambda=0.1$ and  $t_0=8\, \rm  Gyr$ (top
left pannel), $\lambda=0.1$ and  $t_0=3\, \rm Gyr$ (top right pannel),
$\lambda=0.04$  and  $t_0=8\,  \rm   Gyr$  (bottom  left  pannel)  and
$\lambda=0.04$  and  $t_0=3\,  \rm  Gyr$ (bottom  right  pannel),  and
computed  at  different radius  of  the  halo: $r=[5r_s;7r_s]$  (solid
line),  $r=[9r_s;11r_s]$   (dotted  line),  $r=[14r_s;16r_s]$  (dashed
line), $r=[19r_s;21r_s]$ (dash-dotted line).}
\label{windblow}
\end{figure*}

\section{Simulation results}
\label{results}

In this section, we present our simulation results for the 2 different
halo mass considered  in this paper, namely $10^{10}$  and $10^{11} \,
\rm M_{\odot}$.  In  the first case, a strong  wind generally develops
after a  few Gyr. We  will examine in  details the wind  structure and
physical properties. For the second  case, because of the ram pressure
of infalling gas from the surrounding  halo, we see no wind forming: a
clear  galactic  fountain  sets   in  and  creates  a  turbulent,  hot
atmosphere  above (and  below) the  rotating disc.  We will  study the
structure of  this galactic fountain, and compare  the various density
and velocity profiles obtained in both cases (wind versus fountain).

\subsection{$10^{10} \, \rm M_{\odot}$ halo}

In figure \ref{sfr1d10}, we have plotted the star formation rate (SFR)
obtained  in  the   low  mass  galaxies  (Sa,  Sb,   Sc  and  Sd,  see
Table~\ref{tab1} for details). The smallest disc with the highest star
formation efficiency (run Sa) shows,  as expected, the highest peak in
the  star formation  rate. Using  our analytical  model, we  are  in a
position to roughly recover the same curve using $t_* \simeq 1$ Gyr.

Run  Sb and  Sc  have roughly  the  same value  for  the product  $t_0
\lambda_0$, and therefore, as predicted by our simple analytical model
(Eq.~\ref{tstar}-\ref{mgas}),  have roughly  the  same star  formation
history ($t_* \simeq 2$ Gyr).  The largest disc with the lowest star
formation efficency ($t_*  \simeq 4$ Gyr) has the  lowest SFR. At late
time, after  3 to  4 Gyr, the  4 curves  converge to roughly  the same
value.   This corresponds  to the  late epoch  when star  formation is
mainly determined  by the  accretion of gas  infalling from  the outer
halo, as predicted by equation~\ref{mgas}.

Figure \ref{wind1d10_T}  shows a map of  the gas temperature  in a cut
perpendicular  to the  galactic  plane  for run  Sd  ($10^{10} \,  \rm
M_{\odot}$  with $\lambda=0.1$ and  $t_0=8 \,  \rm Gyr$)  at different
epochs.  The  first pannel, at  $t=1.5 \, \rm  Gyr$, shows no  sign of
wind, although it corresponds already to the peak of star formation in
this galaxy (around 0.3 $\, \rm M_{\odot}/\rm yr$). We see several hot
features trying  to break out  of the cold  disc, but none of  them is
able to expand significantly, because of the ram pressure of infalling
halo material.  The second panel corresponds to a later epoch when the
star  formation  rate  has  dropped  by  50\%, down  to  0.2  $\,  \rm
M_{\odot}/\rm yr$  (see also Figure~\ref{wind1d10_dZ}).  Nevertheless,
since gas accretion has also  dropped significantly, two hot and large
cavities now  emerge out  of the  disc and propagate  in the  halo. At
$t=4.5  \, \rm Gyr$  the galactic  wind has  finally escaped  the halo
outer boundaries, and is now in a quasi-permanent regime. This 2 large
cavities have  created some kind of tunnel,  facilitating the ejection
of hot,  supernovae-driven material from  the inner disc, all  the way
out to the  intergalactic medium. At $t=6 \, \rm Gyr$,  we see a final
snapshot of the  galactic wind, with a typical  noozle-like shape (see
Fig.~\ref{wind1d10_T}  and  \ref{wind1d10_dZ}).   The hot,  metal-rich
outflowing  gas is  surrounded  bu  a dense  shell  of compressed  and
cooling  halo gas.   This dense  shell will  eventually  fragment into
small clouds  that will fall back  to the disc.  We  believe that this
dense shell corresponds  to the structure seen in many observations
around M82  (\citealp{heckmanetal90, shopbell&bland98, martin98, hoopesetal03}).   This also corresponds
to the ``superwind'' geometry described in \cite{tenorio&munoz98} (see also \citealp{veilleuxetal05} and references therein).

\subsection{Wind formation epoch}

In order  to describe the  outflow more quantitatively, we  define the
net mass flux across a shell of radius $r_{min}$ and thickness $\Delta r =
r_{max} - r_{min}$ as
\begin{equation}
F=\int_{r_{min}}^{r_{max}} {\rho \vec u . \vec n \over \Delta r} 
4\pi r^2 dr 
= F_+ + F_-\, ,
\label{flux_m}
\end{equation}
where $\vec n=\vec r / r$.  This net flux is further splitted into two
different contribution:  the positive flux,  $F_+$, corresponding only
to outflowing  volume elements,  namely those that  satisfy $\vec  u .
\vec n > 0$, and  the negative flux, $F_-$, corresponding to inflowing
volume elements, satisfying  $\vec u . \vec n < 0$.   We focus here on
the positive flux, in order to detect the outflow, and to estimate the
amount of gas expelled as a  function of time. The shell thickness was
set to $\Delta r  = 2 r_s$, and we vary the  shell radius from $5 r_s$
to $ 20 r_s$,  in order to define a proper radius  where to detect the
outflow. Figure~\ref{windblow}  shows the positive flux  measured at 4
different  radii  for  our 4  low  mass  galaxies,  as a  function  of
time. The  wind is detected when  the positive flux  sharply rise from
zero  to its  maximum value.   The smaller  the shell  radius  is, the
earlier  the  wind  is  detected.   For  our  smallest  shell  radius,
$r_{min}=5  r_s$, the  measured mass  flux remains  quasi  constant in
time, while  for larger  shell radii, the  measured mass  flux sharply
rises to a much larger value, namely 2 to 3 times the inner flux, then
slowly decays to the correct wind mass flux.  This means that when the
wind breaks out  of the halo, a significant fraction (up to $\sim2/3$) of the wind is filled by the hot gaseous halo
and expelled out of the halo boundaries. When the permanent
regime sets  in at later time, the  mass flux is roughly  equal to the
inner  one, independant  of  the  radius of  the  shell. We  therefore
consider  only the  case  $r_{min}=5  r_s$ as  the  correct proxy  for
measuring the mass outflow rate in the wind.

\begin{figure}
\centering{\resizebox*{!}{8.5cm}{\includegraphics{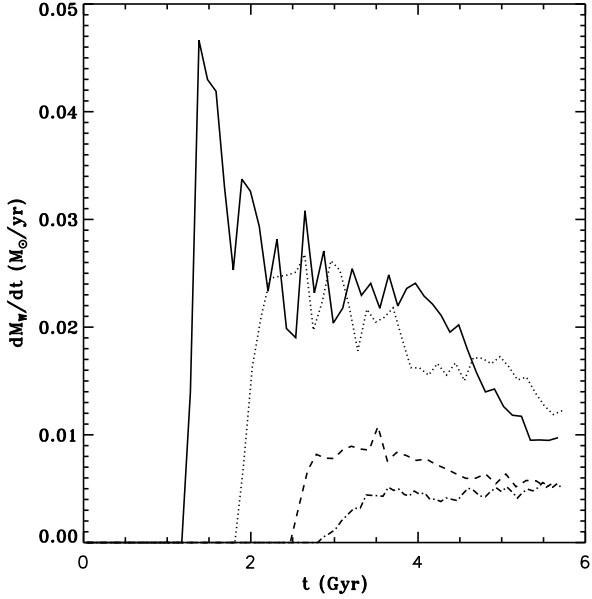}}}
\caption{Mean   flux    of   mass   outflowing    calculated   between
$r=[5r_s;7r_s]$  for  the  $10^{10}   \,  \rm  M_{\odot}$  halos  with
$\lambda=0.04$ and $t_0=3\, \rm  Gyr$ (solid line), $\lambda=0.04$ and
$t_0=8\, \rm  Gyr$ (dotted line), $\lambda=0.1$ and  $t_0=3\, \rm Gyr$
(dashed  line) and  $\lambda=0.1$ and  $t_0=8\, \rm  Gyr$ (dash-dotted
line).}
\label{fluxm}
\end{figure}

We  have  plotted the  positive  flux at  $r_{min}=5  r_s$  for our  4
galaxies   in  figure~\ref{fluxm}.   The   highest  SFR   galaxy,  Sa,
corresponds to our earliest wind formation epoch, around $t_w \simeq 1
Gyr$.  According  to our analytical  model (see figure~\ref{tvsetaw}),
this  corresponds  to  the  hydrodynamical efficency  parameter  $\chi
\simeq 0.9\%$.  Although runs Sb  and Sc have identical star formation
history (see figure~\ref{sfr1d10}), they have different wind formation
epochs, respectively $t_w \simeq$ 2 and 3 Gyr.  This demonstrates that
for a given star formation rate, a more compact disc is more efficient
in forming galactic winds. According  to our analytical recipe, run Sb
corresponds to $\chi \simeq 0.8\%$,  while run Sc barely reaches $\chi
\simeq  0.6\%$. The  final galaxy,  Sd,  features the  latest wind  to
developp, with  $t_w \simeq 4$  Gyr, corresponding also to  a slightly
lower  hydrodynamical   efficency  $\chi  \simeq   0.7\%$.   The  main
conclusion  of  this  section  is  that the  conversion  efficency  of
supernovae energy  deposition rate into wind luminosity  is rather low
in this  model, around  $\chi \simeq 1\%$. This efficiency is directly related on the choice of supernovae energy injection rate, a quite uncertain parameter depending on the scale injection energy  (here chosen equal to $50\%$).   We can see  directly from
figure~\ref{tvsetaw} that, according  to our numerical experiments, no
wind will  ever form in higher  mass halo, as will  be demonstrated in
the next section.

\subsection{Wind efficiency and metallicity}

\begin{figure}
\centering{\resizebox*{!}{8.5cm}{\includegraphics{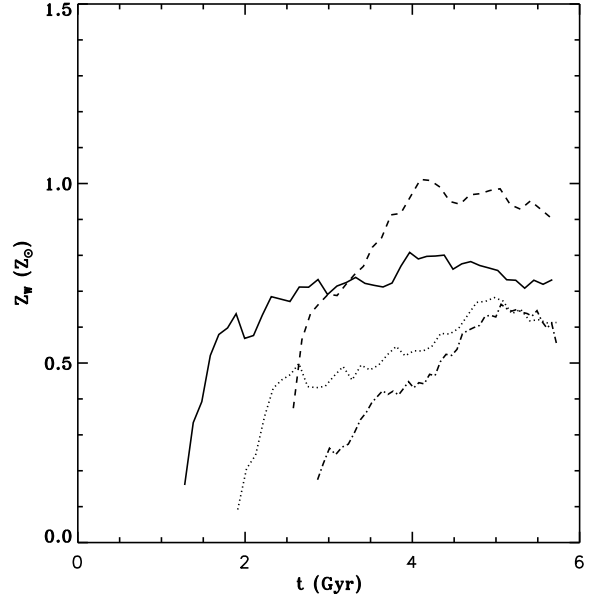}}}
\caption{Mean   metallicity    of   the   wind    calculated   between
$r=[5r_s;7r_s]$  for  the  $10^{10}   \,  \rm  M_{\odot}$  halos  with
$\lambda=0.04$ and $t_0=3\, \rm  Gyr$ (solid line), $\lambda=0.04$ and
$t_0=8\, \rm  Gyr$ (dotted line), $\lambda=0.1$ and  $t_0=3\, \rm Gyr$
(dashed  line) and  $\lambda=0.1$ and  $t_0=8\, \rm  Gyr$ (dash-dotted
line).}
\label{metallicity_wind}
\end{figure}

Before  describing  the  simulation  results for  the  $10^{11}\,  \rm
M_{\odot}$ halo, we would like to characterize the physical properties
of the  galactic winds we  have obtained. It  is common to  define the
wind efficency  as $\eta_w=F_+ / \dot  M_*$. In other  words, the mass
outflow rate is expressed in  units of the global star formation rate.
From  figure~\ref{sfr1d10} and  figure~\ref{windblow}, we  can compute
direclty $\eta_w$:  we find the  maximum efficency at late  time, when
the  permanent regime  is settled,  with values  ranging  from $\eta_w
\simeq  0.05$ for Sc  and Sd,  up to  $\eta_w \simeq  0.1$ for  Sa and
Sb. Here again,  it appears clearly that more  compact discs (low spin
parameter  $\lambda_0$)  gives more  efficient  winds, while  extended
discs (high spin parameter) are  less efficient.  All these values are
extremely  low   compared  to   the  wind  efficiencies   observed  by
\cite{martin99} in  Lyman break galaxies  with values greater  than 1.
These high-redshift galaxies are  likely massive starbursts, for which
our quiescent  approach does  not apply.  On  the other hand,  we have
determined  that compact star  formation sites  result in  earlier and
stronger winds.  Nuclear starbursts  can be natural candidates to host
very  efficient  winds with  $\eta_w  \ge  1$,  although these  bright
galaxies are probably rather rare objects in the early universe.

As seen from figure~\ref{wind1d10_dZ}, the hot gas carried away in the
wind is highly  enriched with metals ($Z_w \simeq  \, \rm Z_{\odot}$).
We define the average wind  metallicity as the positive flux of metals
divided  by  the positive  flux  of mass  $Z_w=(\rho  Z  u)_+ /  F_+$.
Figure~\ref{metallicity_wind} shows  the mean metallicity  of the wind
for our various  low mass galaxies. We see no clear  trend of the wind
metallicity as a  function of disc size or  star formation efficiency.
We can only  observe that the metallicity of the wind  is of about 0.5
to 1 $\rm  Z_{\odot} $. This rather high value  confirms that the wind
comes mainly  from the gas within  the galactic disc  that is directly
enriched by exploding supernovae.

We  now briefly  discuss how  these winds  might explain  the observed
metallicity  of the  intergalactic  medium (IGM).   We  know from  our
simulations that  the typical  mass outflow rate  is around  $\dot M_w
\simeq 0.01 \, \rm M_{\odot}/\rm yr$ (see figure~\ref{fluxm}), with an
average  wind metallicity of  $Z_w \simeq  1 \,  \rm Z_{\odot}$  and a
typical   wind   velocity   around   $u_w  \simeq   300$   km/s   (see
figure~\ref{vprofiles}). If we assume that the IGM is photo-ionized to
a temperature  around $10^4$-$10^5$ K,  we can compute the  volume $V$
occupied  by  the  wind  when  the  bubble  stalls  by
pressure equilibrium with the IGM as
\begin{equation}
\dot M_w u_w^2 t \simeq {P_{IGM} V \over \gamma -1}\, ,
\end{equation}
assuming  the gas  is  adiabatic.  After  6 Gyr,  the
radius $R  \simeq V^{1/3}$  of the expanding  wind lies around  1 Mpc.
This corresponds to  a swept up mass of  baryons roughly $5\times 10^9
\, \rm  M_{\odot}$.  If we assume  that the metals carried  out by the
wind perfectly  mix with the  zero-metallicity IGM, we obtain  for the
metallicity of the IGM the following very simple relation
\begin{equation}
Z_{IGM} \simeq \left( \frac{c_s}{u_w} \right)^2 \,  
\rm Z_{\odot} \simeq 10^{-3}-10^{-2}  \,  \rm Z_{\odot}
\end{equation}
where $c_s \simeq$ 10-30 km/s is the sound speed of the IGM. Note that
this analysis is valid only  for the idealized case we have considered
here.  Observations of the $\rm Ly\alpha$ forest clouds show that they
are  metal enriched  from  $Z  \simeq 10^{-3}$  up  to $10^{-2}\,  \rm
Z_{\odot}$    (\citealp{songaila&cowie96},   \citealp{ellisonetal00}).
Cosmological simulations also confirm this trend, although the various
wind models  proposed (including our) might be  difficult to reconcile
(\citealp{gnedin98}, \citealp{thackeretal02}).

\subsection{$10^{11}\, \rm M_{\odot}$ halo}

\begin{figure}
\centering{\resizebox*{!}{8.5cm}{\includegraphics{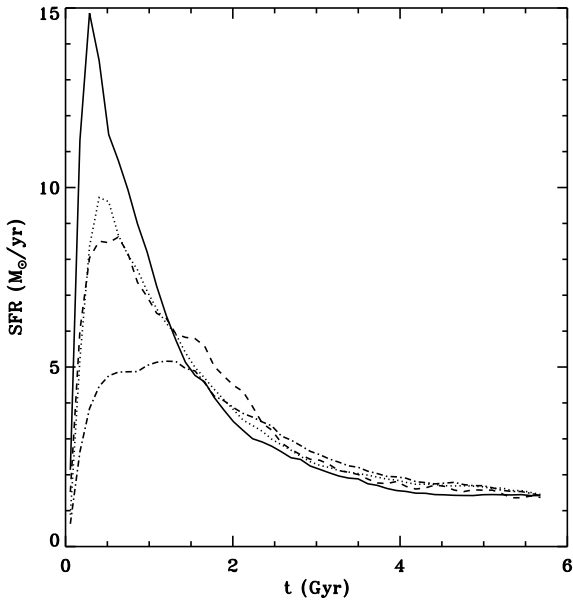}}}
\caption{Star formation rates of  the $10^{11} \, \rm M_{\odot}$ halos
with $\lambda=0.04$ and $t_0=3\, \rm Gyr$ (solid line), $\lambda=0.04$
and $t_0=8\,  \rm Gyr$ (dotted  line), $\lambda=0.1$ and  $t_0=3\, \rm
Gyr$ (dashed  line) and $\lambda=0.1$ and $t_0=8\,  \rm Gyr$ (dash-dotted
line).}
\label{sfr1d11}
\end{figure}

\begin{figure*}
\centering{\resizebox*{!}{8.5cm}{\includegraphics{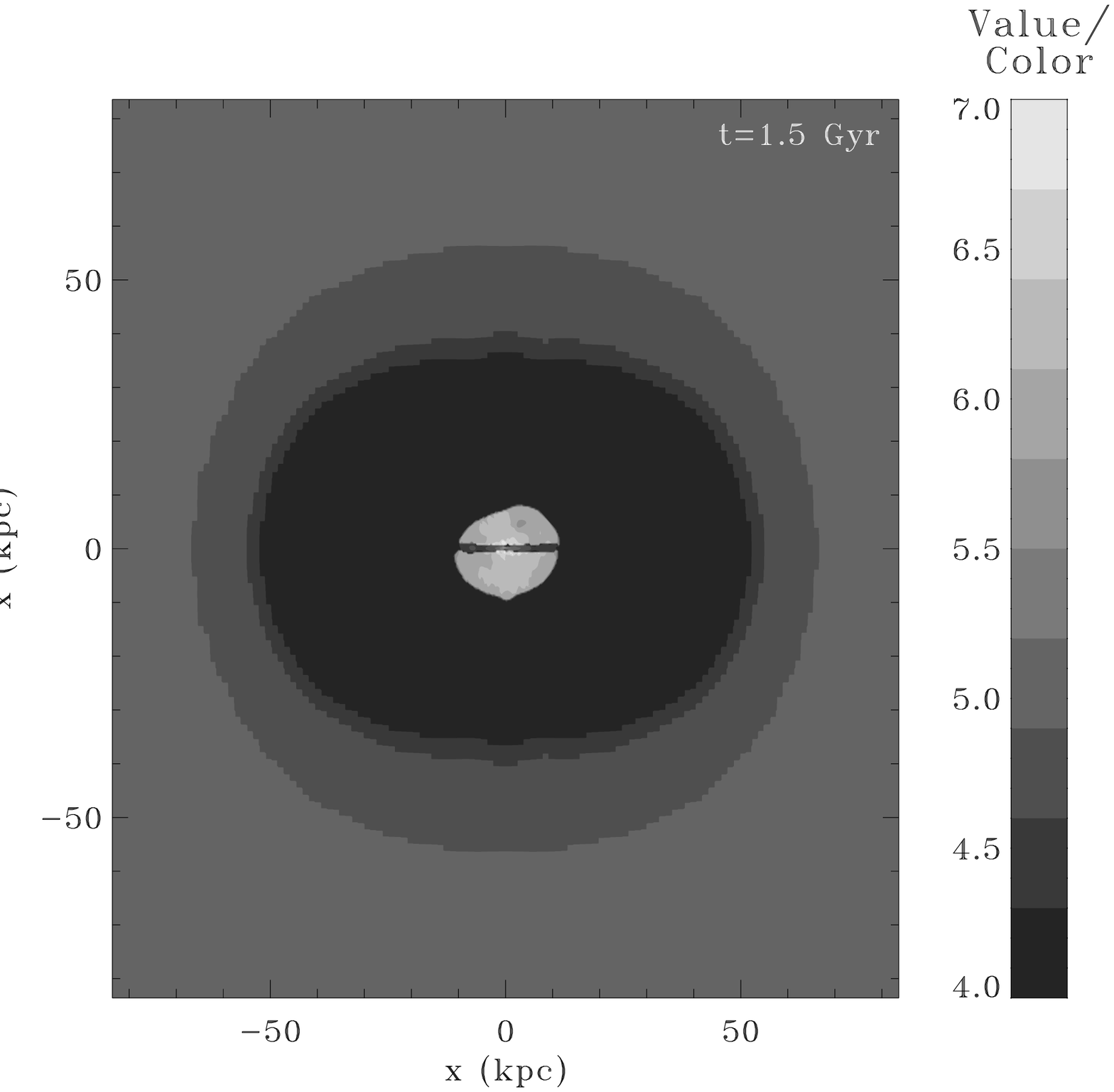}}}
\centering{\resizebox*{!}{8.5cm}{\includegraphics{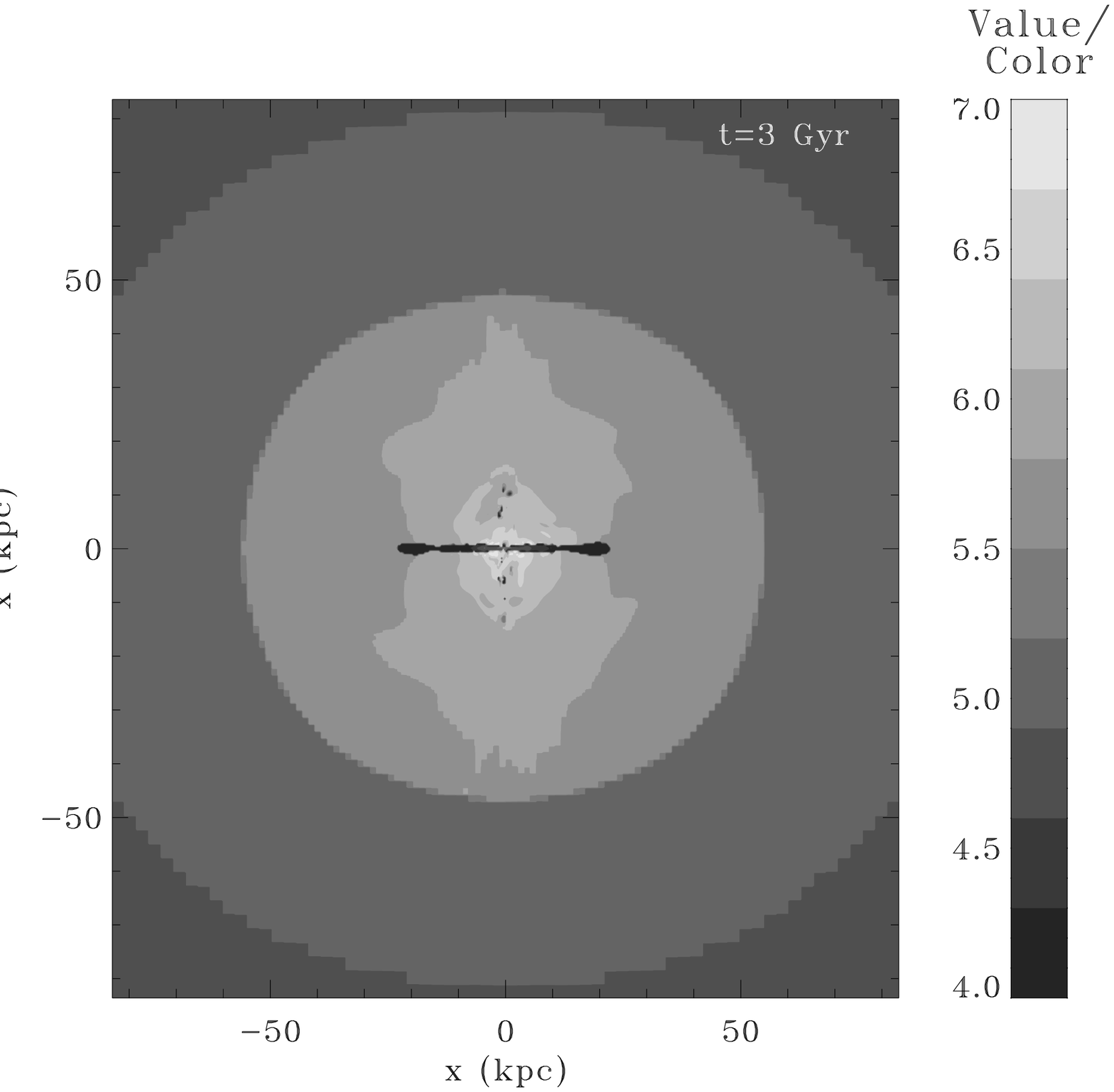}}}
\centering{\resizebox*{!}{8.5cm}{\includegraphics{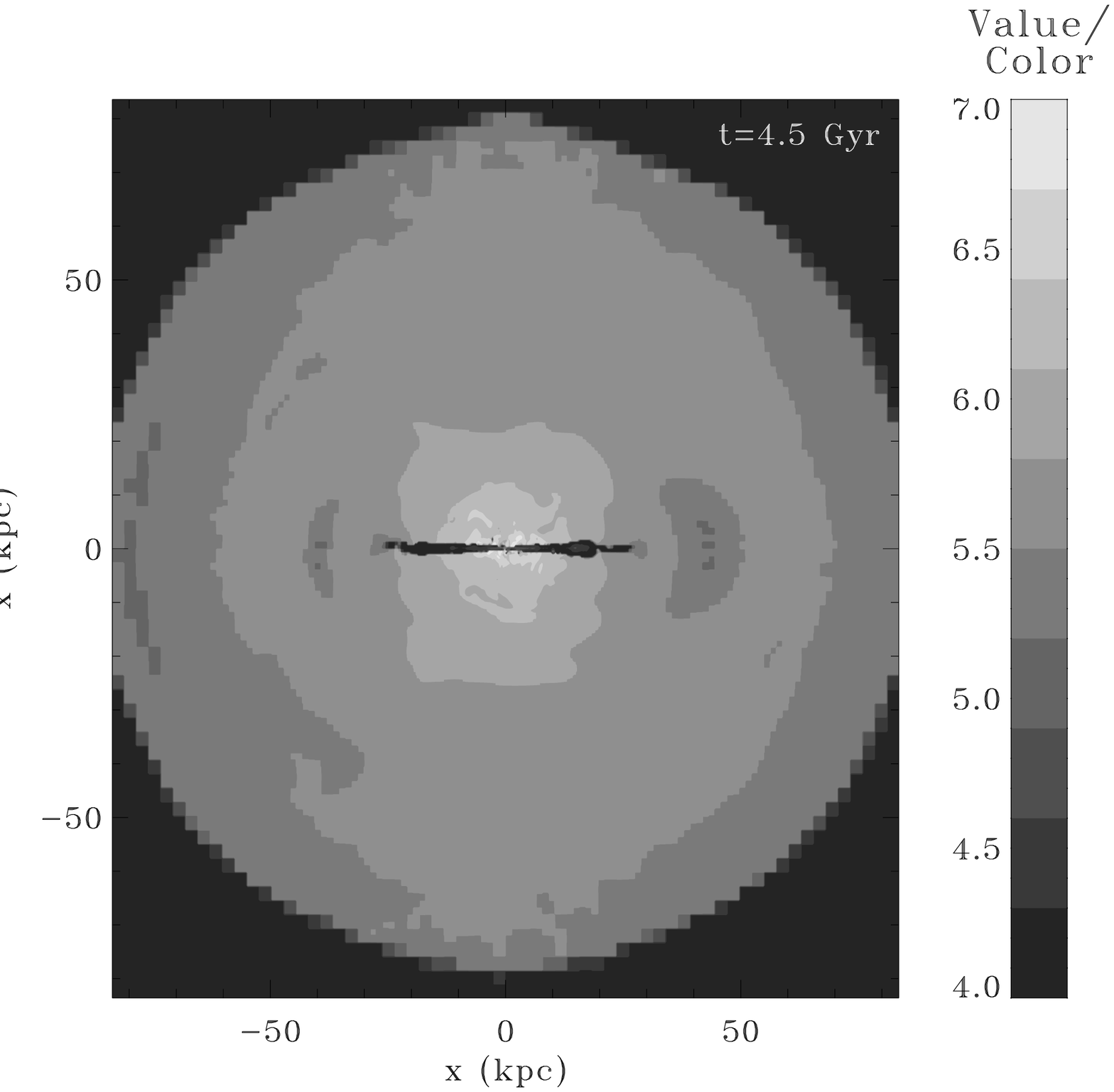}}}
\centering{\resizebox*{!}{8.5cm}{\includegraphics{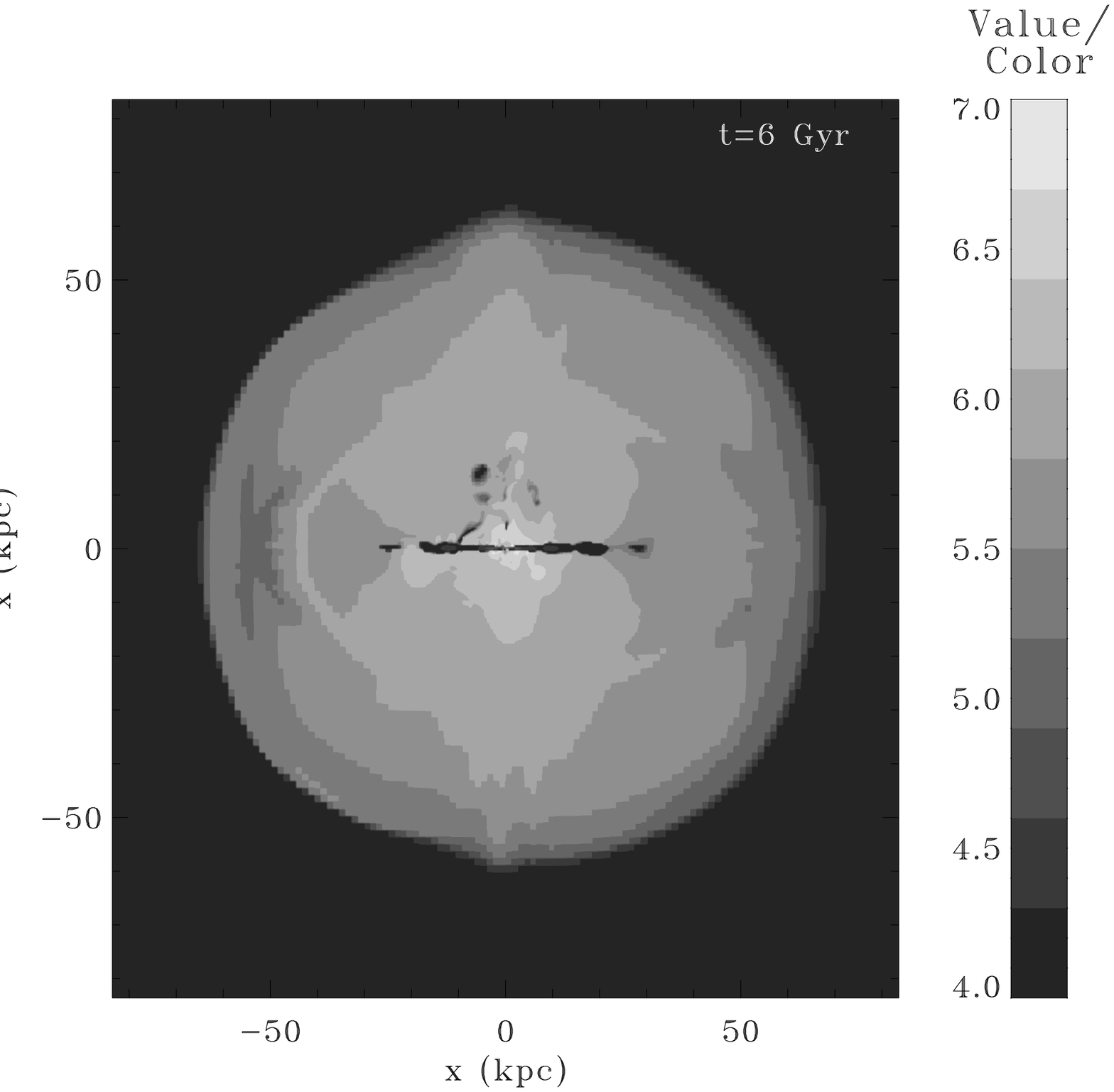}}}
\caption{Cut  of  the  gas  temperature  in  the  Oyz  plane  for  the Ld 
simulation at different epochs. The 4 pannels
are a  zoom 4 times  of the simulation  box. The colour scale  gives the
temperature in $\log (\rm K)$.  Note that length scales are not the same in each pannel.}
\label{wind1d11_T}
\end{figure*}

\begin{figure*}
\centering{\resizebox*{!}{8.5cm}{\includegraphics{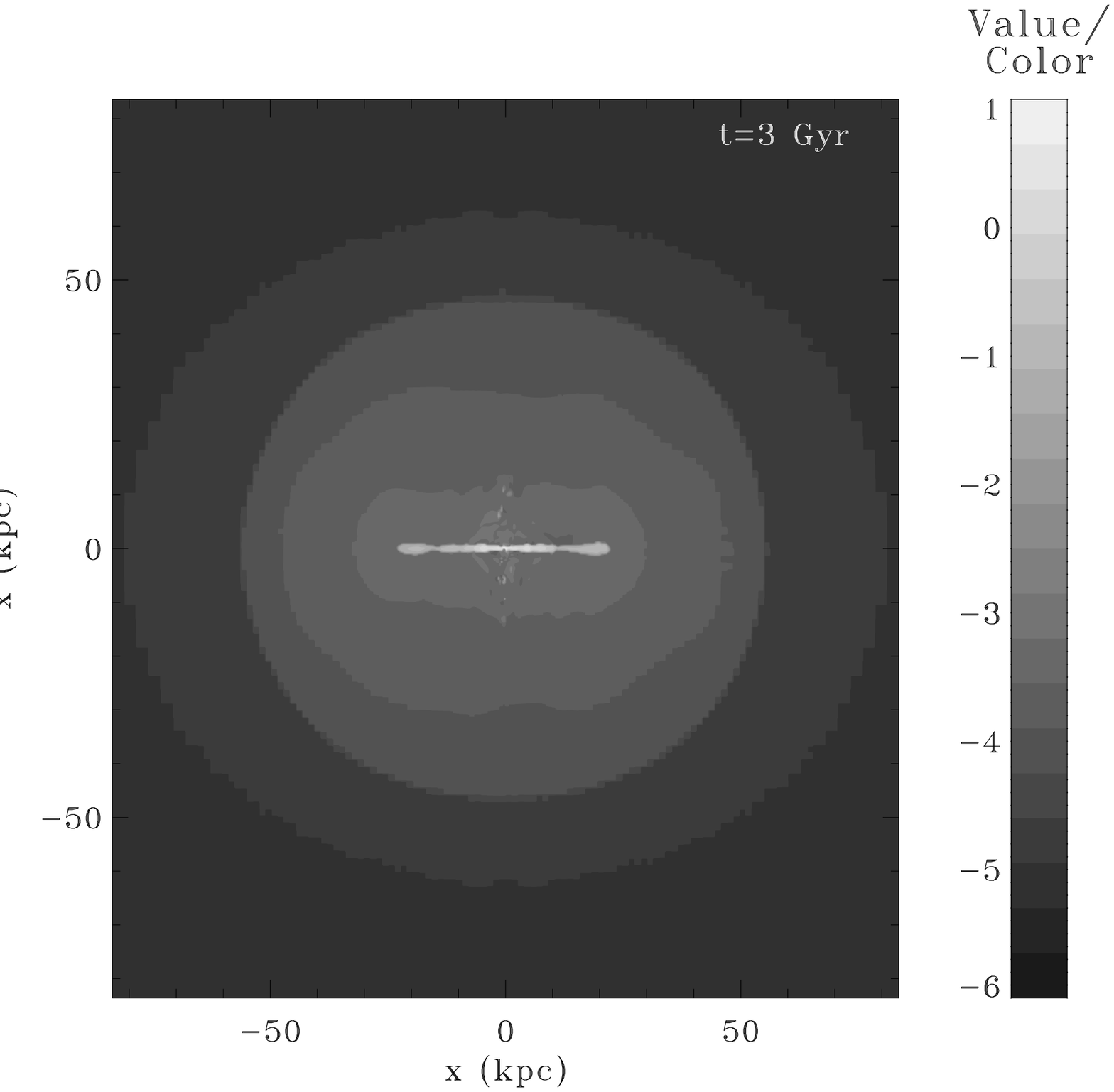}}}
\centering{\resizebox*{!}{8.5cm}{\includegraphics{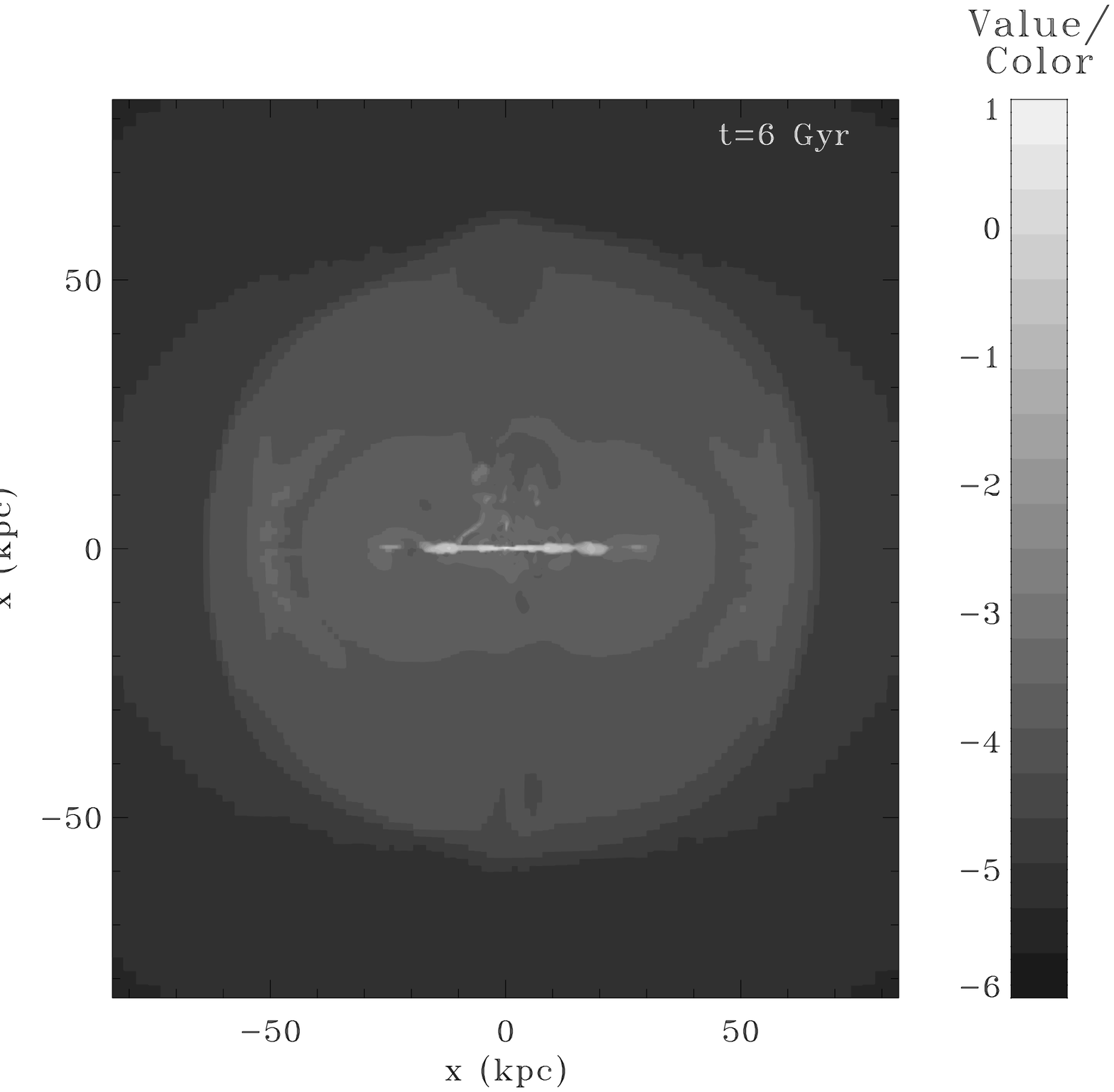}}}
\centering{\resizebox*{!}{8.5cm}{\includegraphics{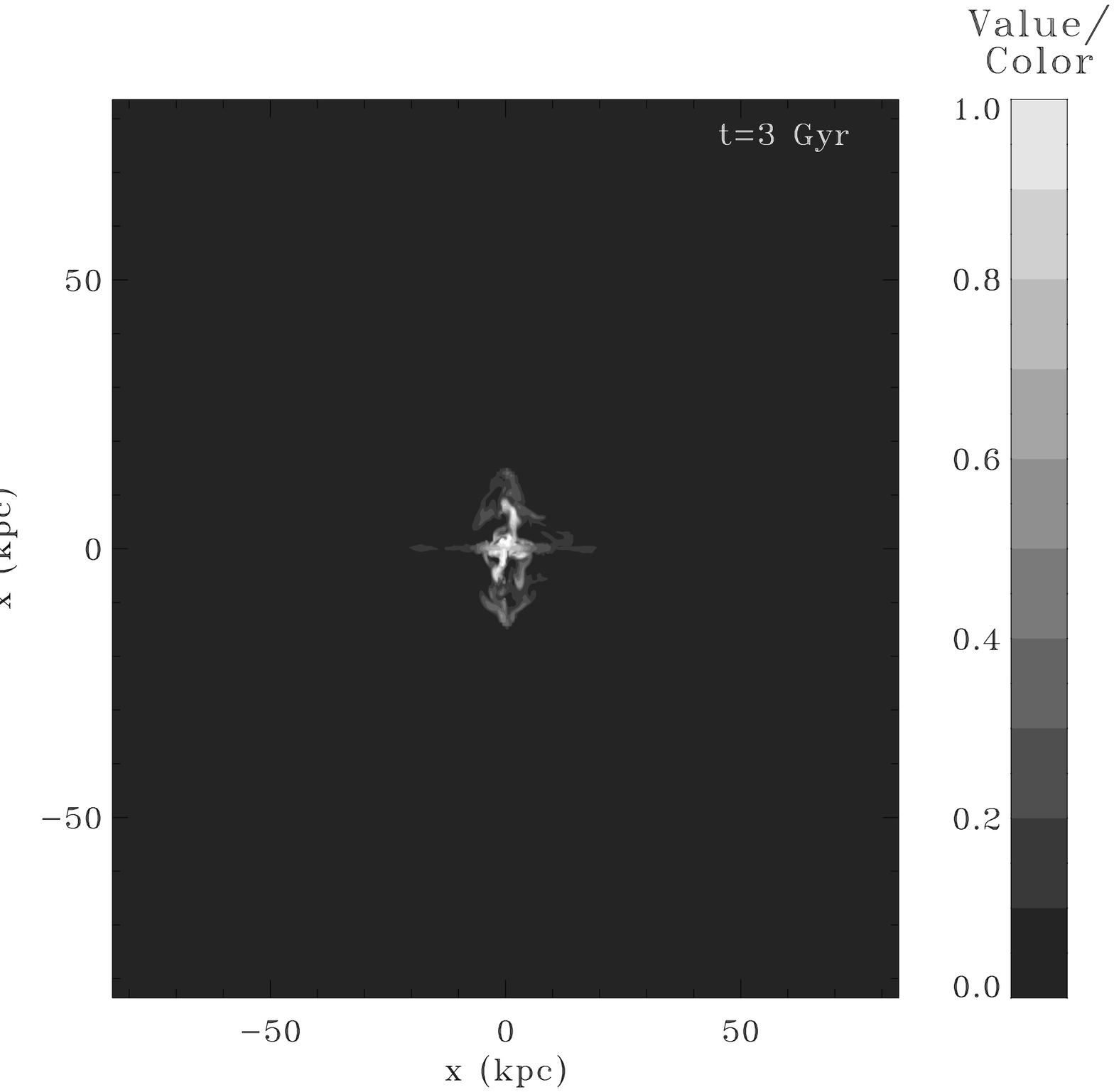}}}
\centering{\resizebox*{!}{8.5cm}{\includegraphics{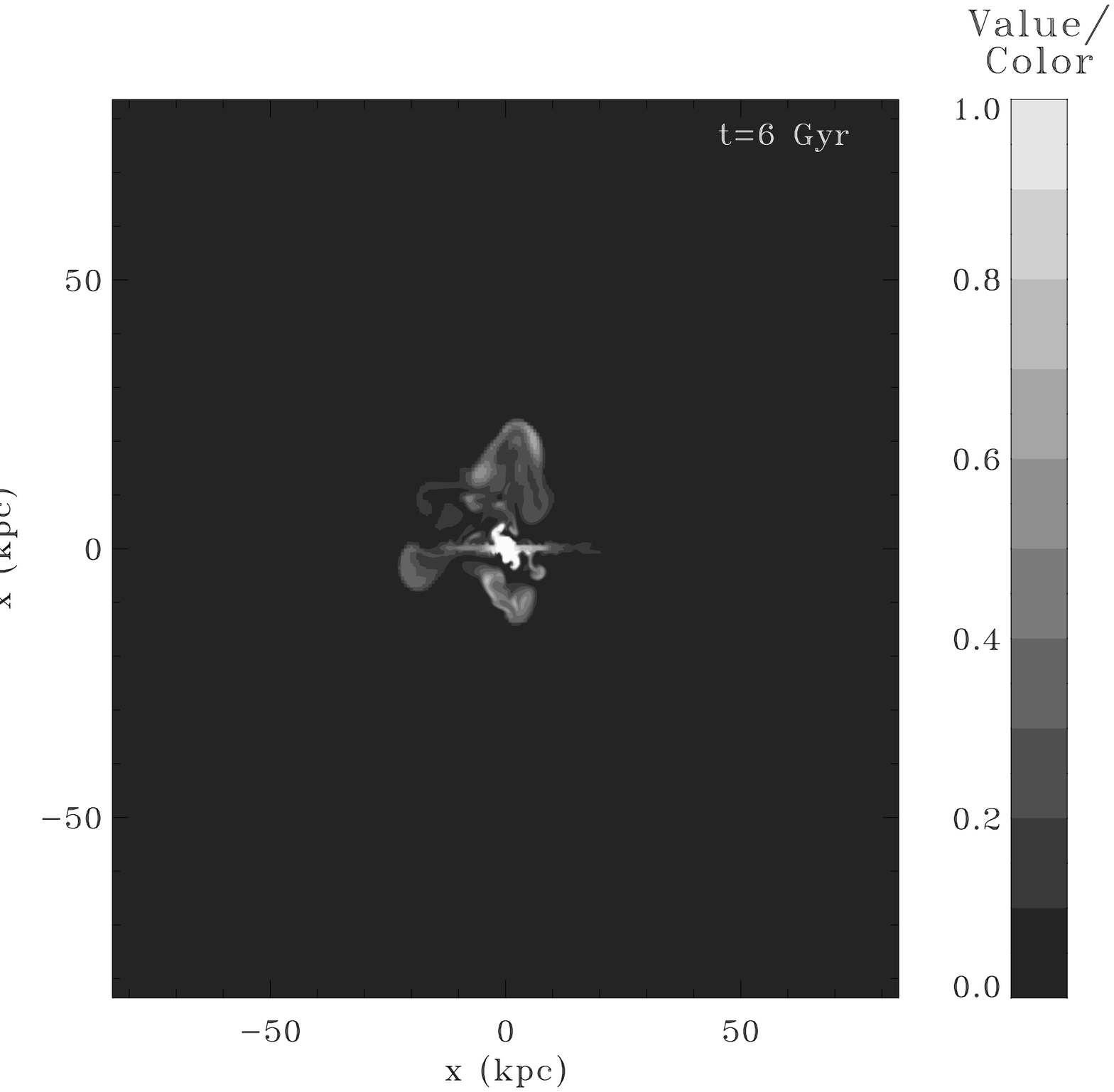}}}
\caption{Cut  of  the  gas density  (up)  and metallicity (bottom) in  the  Oyz plane  for  the 
Ld simulation at different epochs. The two left
pannels are  a zoom 4  times of the  simulation box and the  two right
pannels  are the  entire  simulation  box. The  colour  scale gives  the
density  in $\log(\rm  {cm^{-3}})$ (up)  and the  metallicity  in $\rm
Z_{\odot}$ (bottom). Note that length scales are not the same in each pannel.}
\label{wind1d11_dZ}
\end{figure*}

We have plotted in figure \ref{sfr1d11} the star formation history for
the 4 galaxies hosted by the $10^{11}\, \rm M_{\odot}$ halo.  They are
qualitatively very similar to the $10^{10}\, \rm M_{\odot}$ case, with
La  and Ld  as the  most extreme  models, and  Lb and  Lc  being quasi
undistinguishable.  Each galaxy has however a peak star formation rate
which  is 50\%  higher  than its  rescaled  $10^{10}\, \rm  M_{\odot}$
counterpart.  The  circular velocity in  these new galaxies  is higher
than the  low mass  ones, while  the gas sound  speed, because  of the
polytropic multiphase model, remains roughly constant, around 10~km/s,
so that  the disc is  now much thinner.  This results in  higher total
star  formation  efficencies.  Using  our  analytical  model,  we  can
reproduce the 3 star formation histories using for run Sa, Sb (and Sc)
and Sd the values $t_* \simeq$ 0.75, 1.5 and 3 Gyr respectively.

The other  striking differences  with low mass  galaxies is  the clear
presence of an  accretion shock surrounding the disc,  and the notable
absence  of  galactic  winds.   We  observe instead  a  nice  galactic
fountain, with plumes  of hot gas rising above  the disc, cooling down
and falling back to the disc (see figure~\ref{wind1d11_T}). 

\begin{figure*}
\centering{\resizebox*{!}{8cm}{\includegraphics{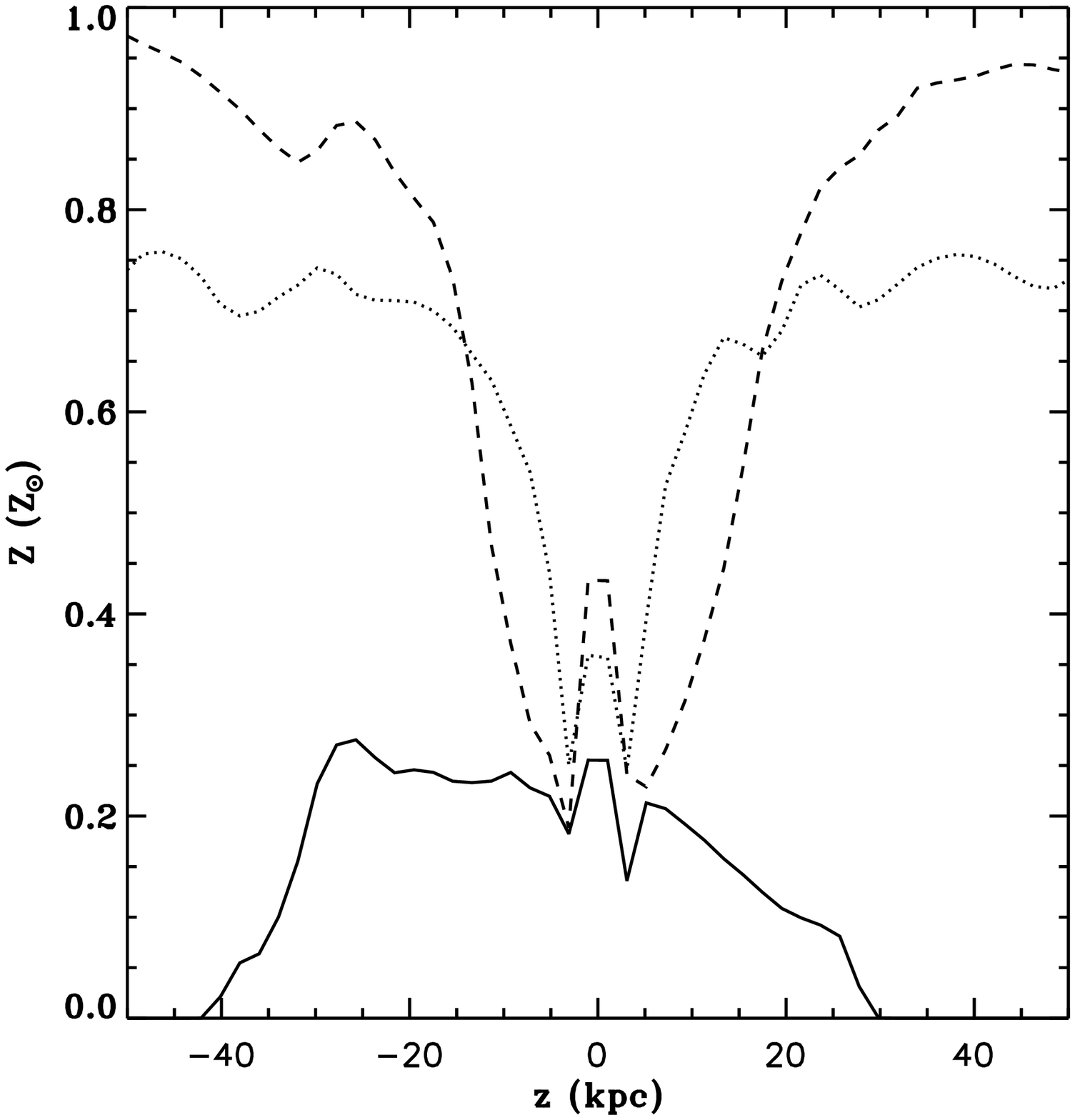}}}
\centering{\resizebox*{!}{8cm}{\includegraphics{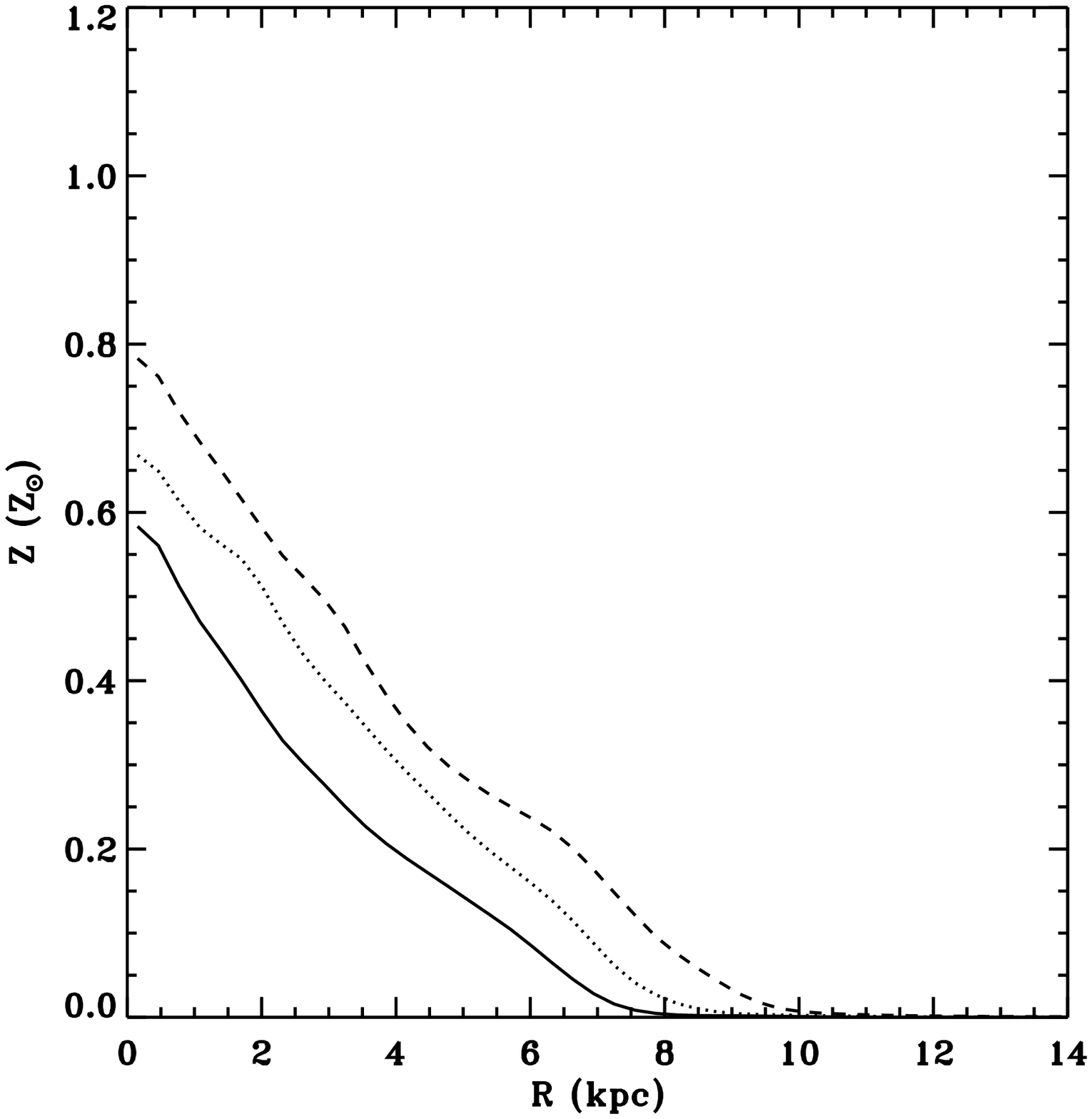}}}
\centering{\resizebox*{!}{8cm}{\includegraphics{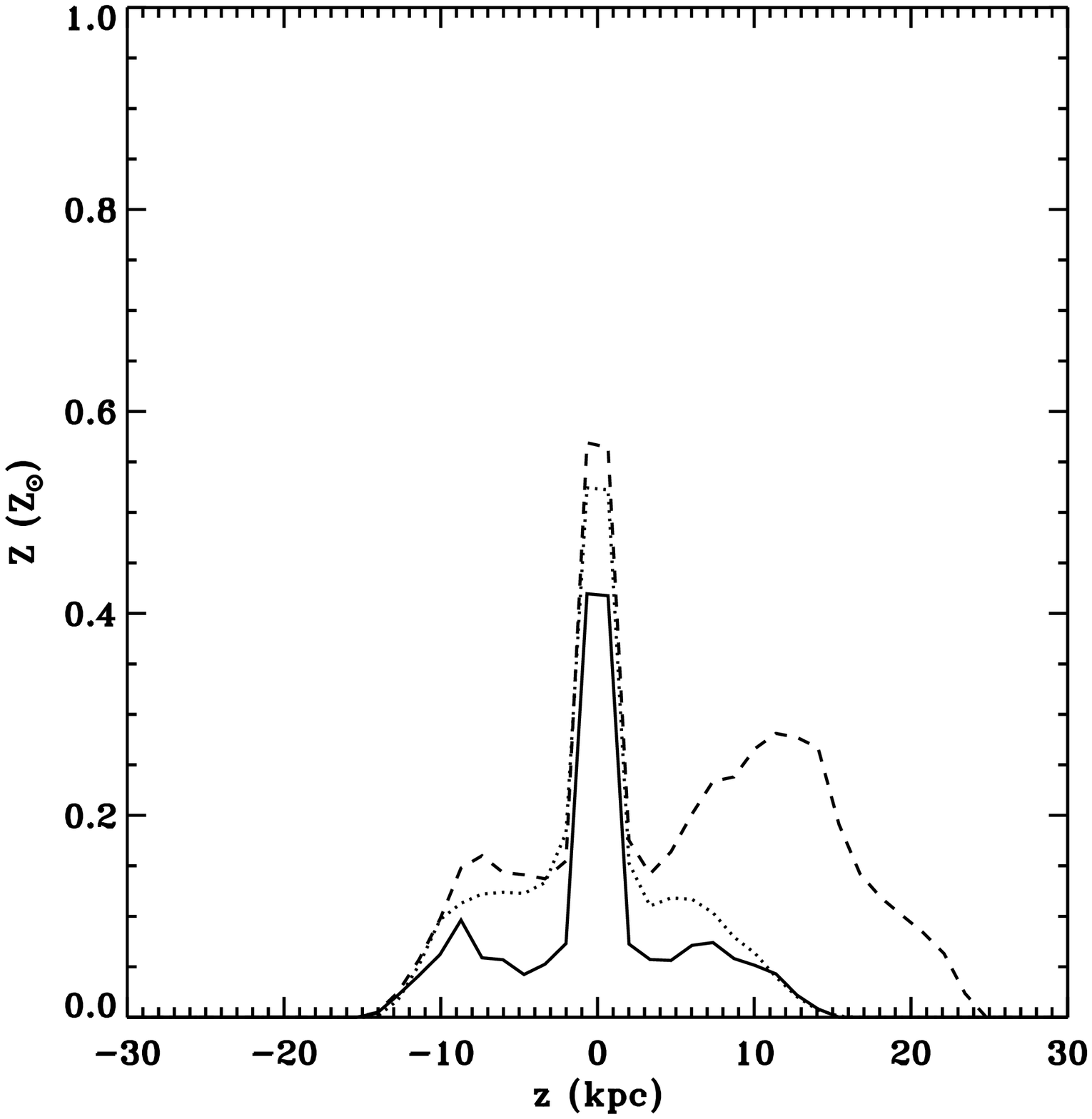}}}
\centering{\resizebox*{!}{8cm}{\includegraphics{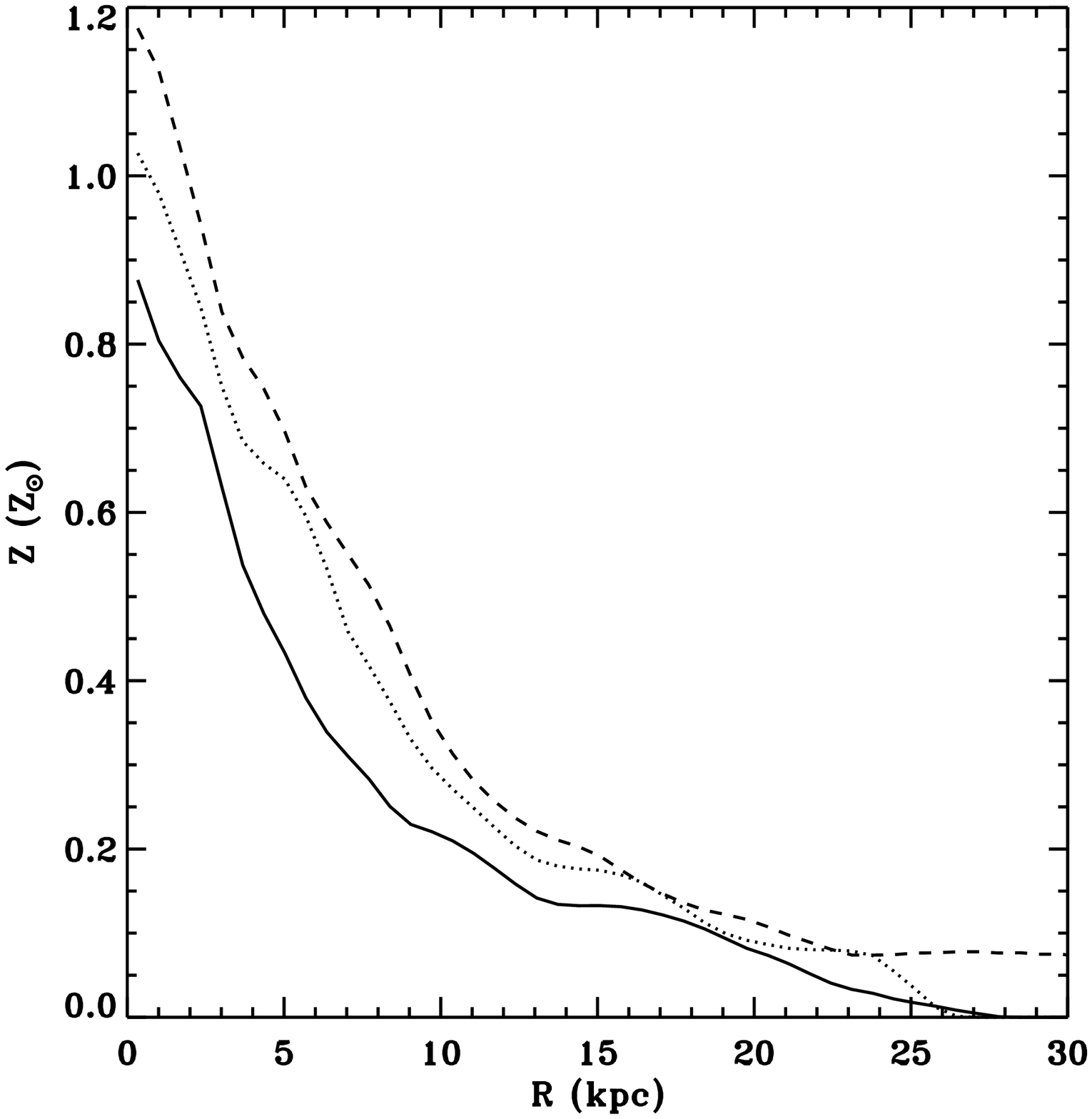}}}
\caption{Mean metallicity  of the gas  as a function of  the projected
height (left)  and radius (right)  for Sd simulation (up)  and  the  Ld simulation  (bottom)  at
different times:  $t=3 \,  \rm Gyr$ (solid  line), $t=4.5 \,  \rm Gyr$
(dotted line), $t=6 \, \rm Gyr$ (dashed line).}
\label{Zprofiles}
\end{figure*}

The formation of  an accretion shock above a certain  halo mass is now
well  understood: the stability  analysis of  a radiative  shock wave,
performed by  \cite{birnboim&dekel03}, explains why for  low mass halo
no accretion  shock forms.  The  mass threshold above which  the shock
forms  depends on  the metallicity  of the  cooling gas.   We  see from
figure~\ref{wind1d11_T} that  in our  simulation the shock  heated gas
above   and  below   the   disc   is  metal   poor:   in  this   case,
\cite{birnboim&dekel03}  found a  critical mass  for  shock appearance
slightly below  $10^{11}\, \rm  M_{\odot}$ for a zero metallicity gas, in good  agreement with
our  numerical  experiment.   The  accretion  shock
converts the infall  kinetic energy into internal energy,  so that our
analytical treatment  of wind  formation remains valid,  replacing the
ram pressure by  the thermal pressure. The accretion  shocks reaches a
maximum radius of $\sim 75 \,  \rm kpc$ what is $\sim 2/3 \, r_{vir}$ quite the same virial radius fraction than the one found in \cite{birnboim&dekel03} for the cooling case after $\sim4-5$ Gyr, and finally
contracts down  to $\sim 50 \,  \rm kpc$ at later  time.  Although the
disc  is strongly  perturbed by  supernovae blast  waves,  no galactic
winds can form, in agreement with our simple analytical prediction, if
we assume for the hydrodynamical  efficency $\chi \simeq 1\%$, a value
close to the one obtained in the $10^{10}\, \rm M_{\odot}$ case.  The
atmosphere above  and below the  disc is however highly  turbulent and
perturbed by  buyoantly rising plumes.   This convective flow  is more
visible in the metallicity map of figure~\ref{wind1d11_dZ}.

Metals are  confined within  a rather small  distance relative  to the
disc plane,  namely $\sim 20  \, \rm kpc$.  In this case,  the thermal
pressure of the shock--heated halo  gas is very efficient in confining
the  galactic wind.  Figure  \ref{Zprofiles} compares  the metallicity
profiles of  the gas between the  $10^{10} \, \rm  M_{\odot}$ halo and
the  $10^{11}  \,  \rm M_{\odot}$  halo  (run  Sd  versus run  Ld)  at
different epochs.  We define  for that purpose a mass-weighted average
metallicity as a function of height and a mass-weighted metallicity 
as a function of radius as
\begin{equation}
Z(z)=\frac{\int_0^{r_d} \rho Z 2\pi r dr }
{\int_0^{r_d} \rho 2\pi r dr }~~~{\rm and}~~
Z(r)= \frac{\int_{-h_d}^{+h_d} \rho Z 2\pi dz} 
{\int_{-h_d}^{+h_d} \rho 2\pi dz}
\end{equation} 
where $\rho$  is the gas density  and $Z$ is the  gas metallicity. The
disc radius  $r_d$ is taken equal  to $r_s$ and the  disc scale height
$h_d=r_d/10$.  In  the wind case,  metals are entrained at  very large
distance above and below the disc in the wind noozle.  The metallicity
is roughly solar and tends to  saturate at that value. In the fountain
case,  the metallicity  in the  halo remains  rather low,  roughly one
tenth solar.   On the other hand,  the gas metallicity  in the central
part  of  the disc  is  significantly higher  than  in  the wind  case
($Z_{gas} \simeq 1.2  \, \rm Z_{\odot}$ versus $Z_{gas}  \simeq 0.6 \,
\rm Z_{\odot}$, see figure~\ref{Zprofiles}). This could be interpreted
as  the effect  of the  wind removing  metals preferentially  from the
bulge of the galaxy. In the  outer parts of the disc, however, the gas
metallicity  is  very  similar  in  both cases.  This  could  also  be
interpreted as  a higher  star formation efficiency  in the  high mass
case, especially in the central parts where the density is higher.  As
explained  in  great  details   in  \cite{dalcanton06},  it  is  quite
difficult to  disentangle the influence of star  formation, gas infall
and winds on  metal enrichment in galaxies.  We  believe that, even in
our idealized case, the metal content in our galaxies is determined by
a subtle balance between these 3 processes.

\subsection{Stellar distribution and metallicity} 

\begin{figure*}
\centering{\resizebox*{!}{8cm}{\includegraphics{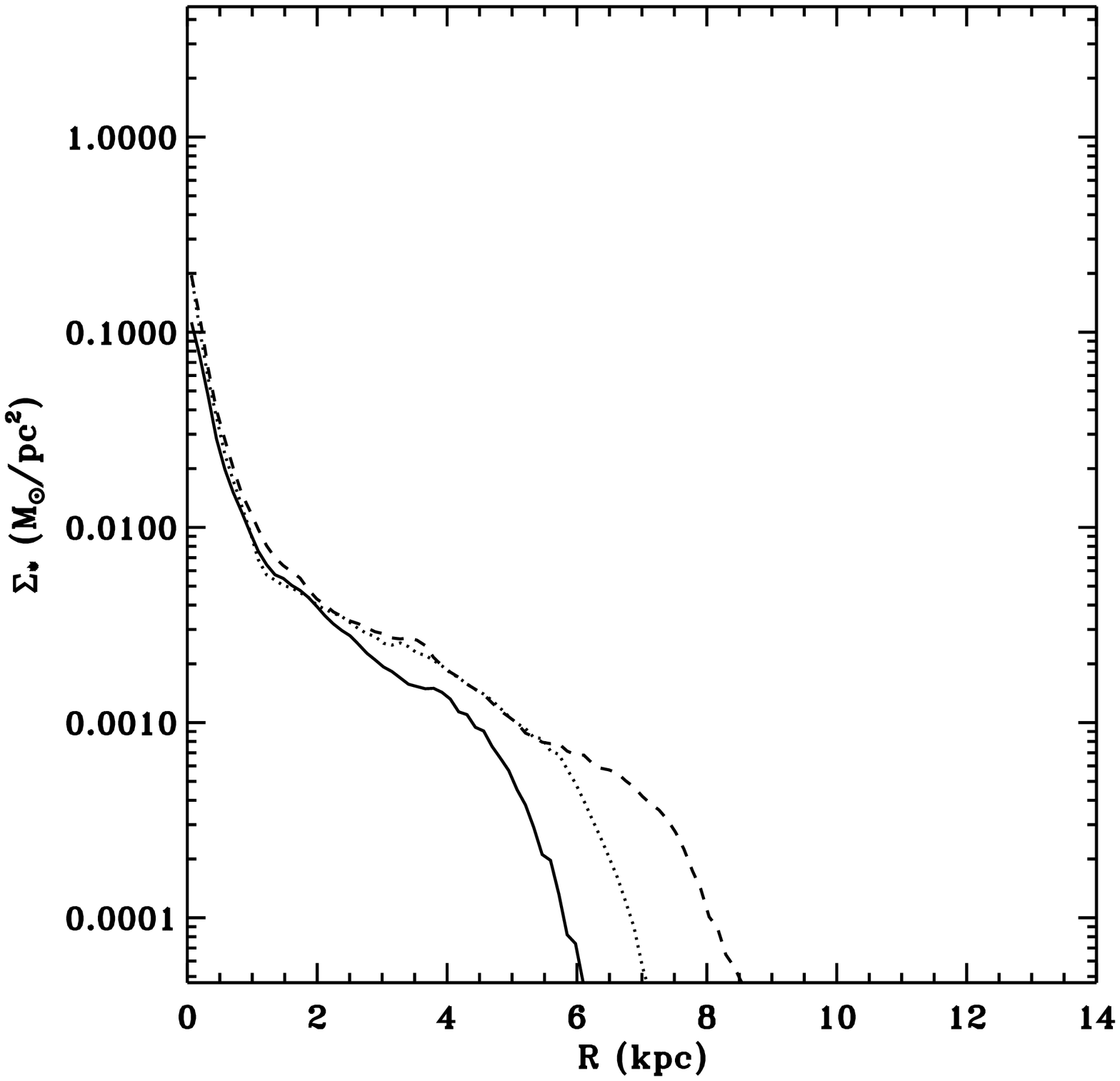}}}
\centering{\resizebox*{!}{8cm}{\includegraphics{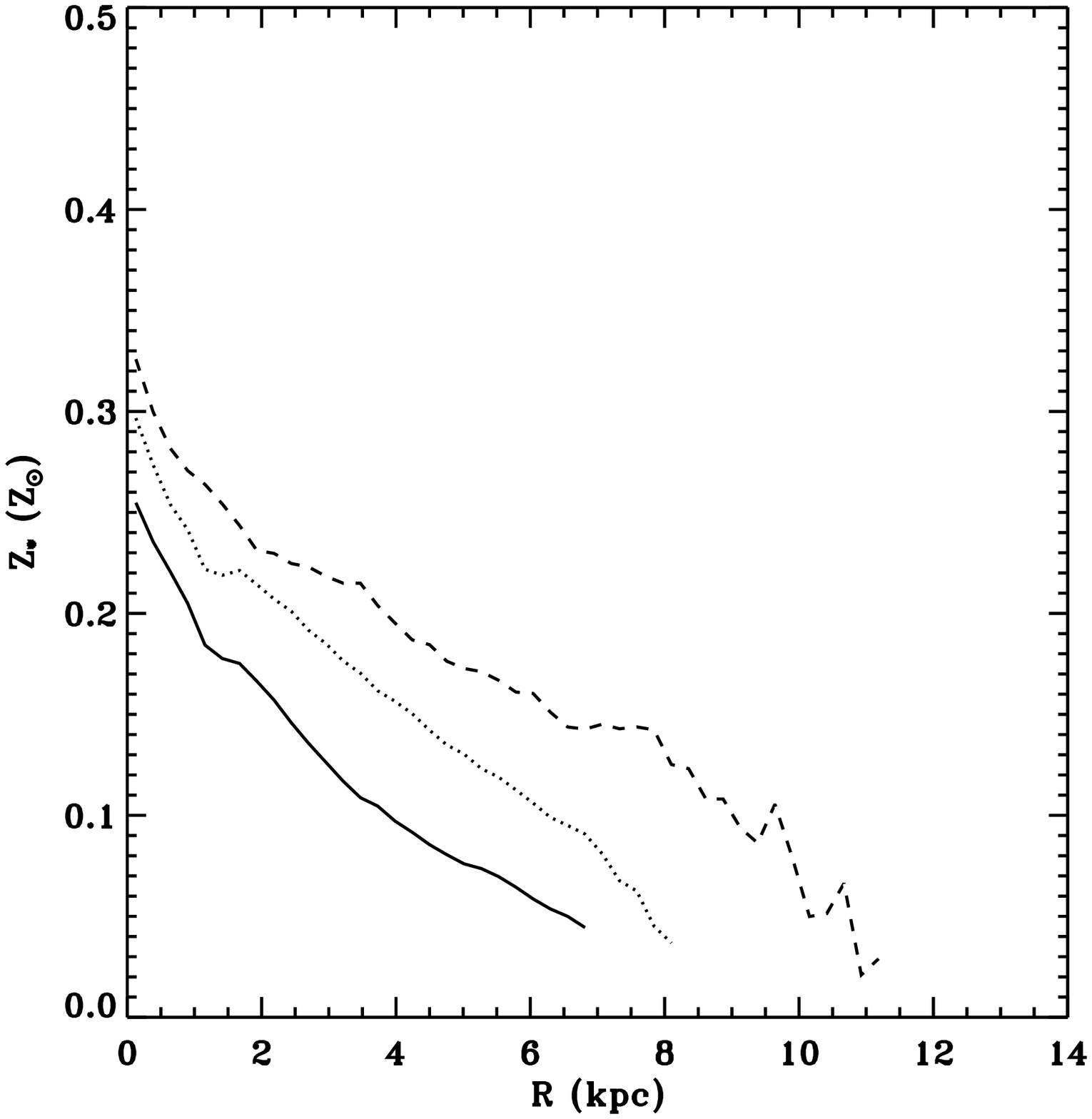}}}
\centering{\resizebox*{!}{8cm}{\includegraphics{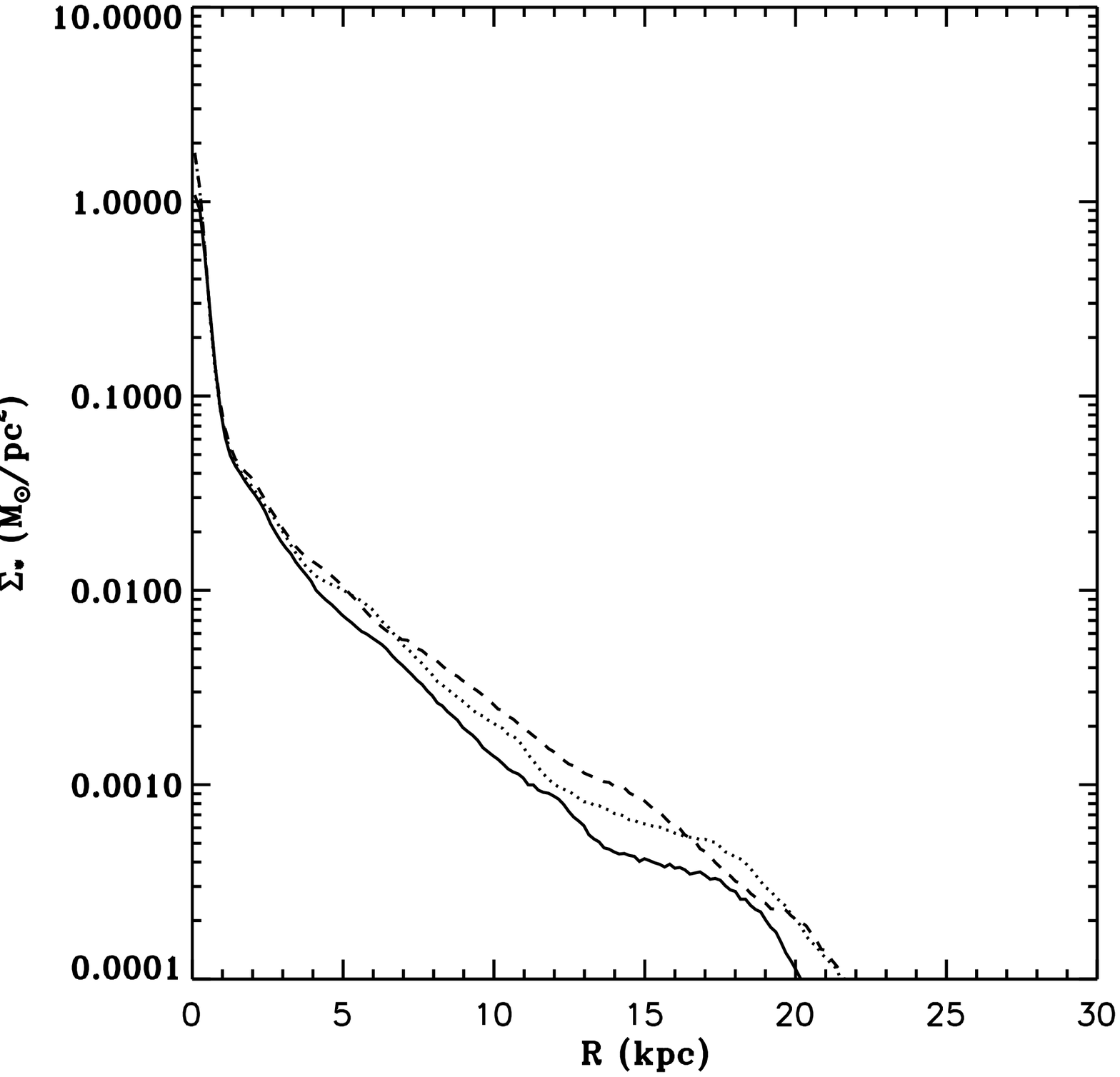}}}
\centering{\resizebox*{!}{8cm}{\includegraphics{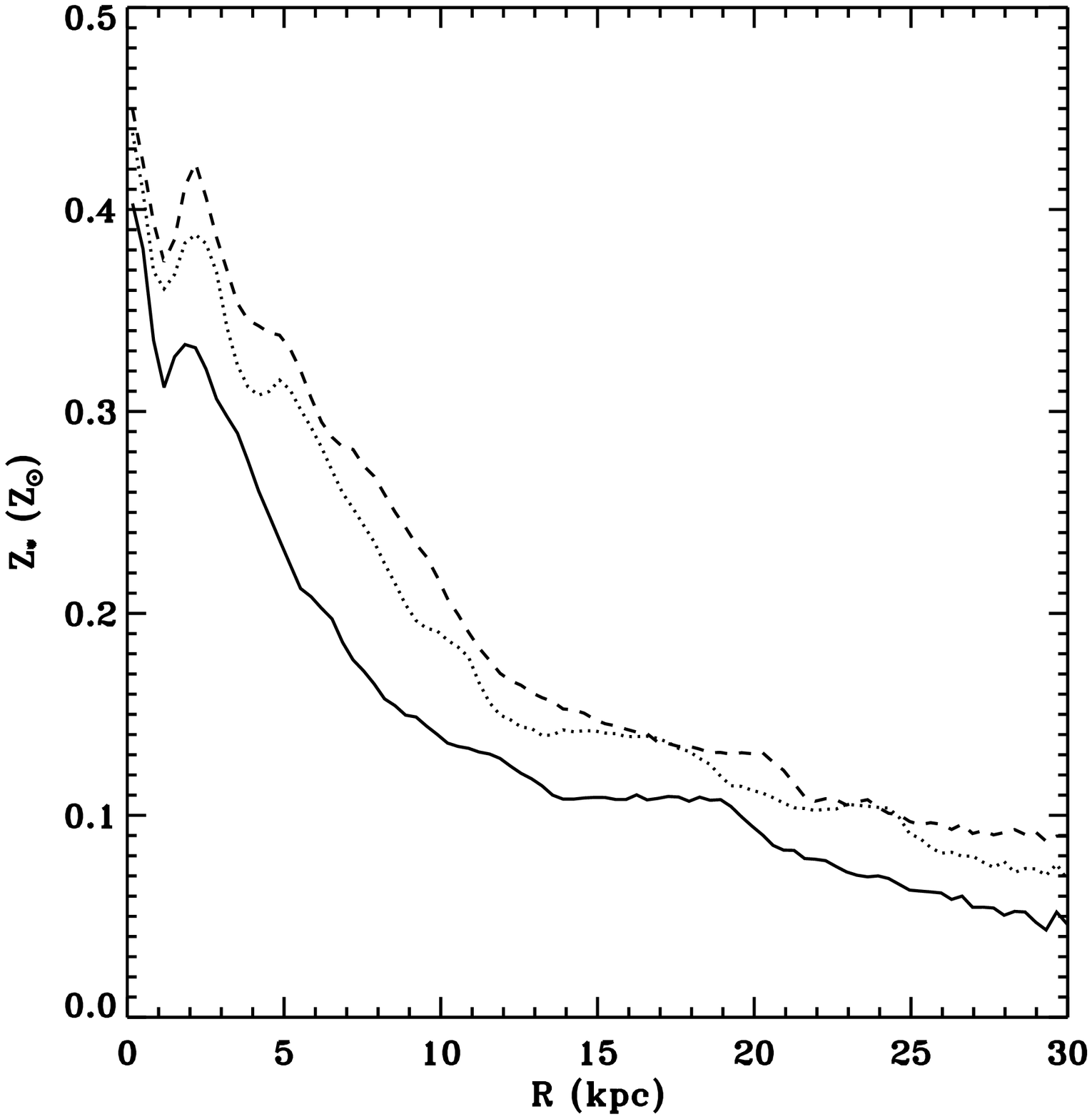}}}
\caption{Surface density  (left) and  mean metallicity (right)  of the
stars as  a function of the  projected radius for the  Sd simulation (up) and the Ld simulation (bottom)
at different times: $t=3 \, \rm  Gyr$ (solid line), $t=4.5 \, \rm Gyr$
(dotted line), $t=6 \, \rm Gyr$ (dashed line).}
\label{starsprofiles}
\end{figure*}

\begin{figure*}
\centering{\resizebox*{!}{7cm}{\includegraphics{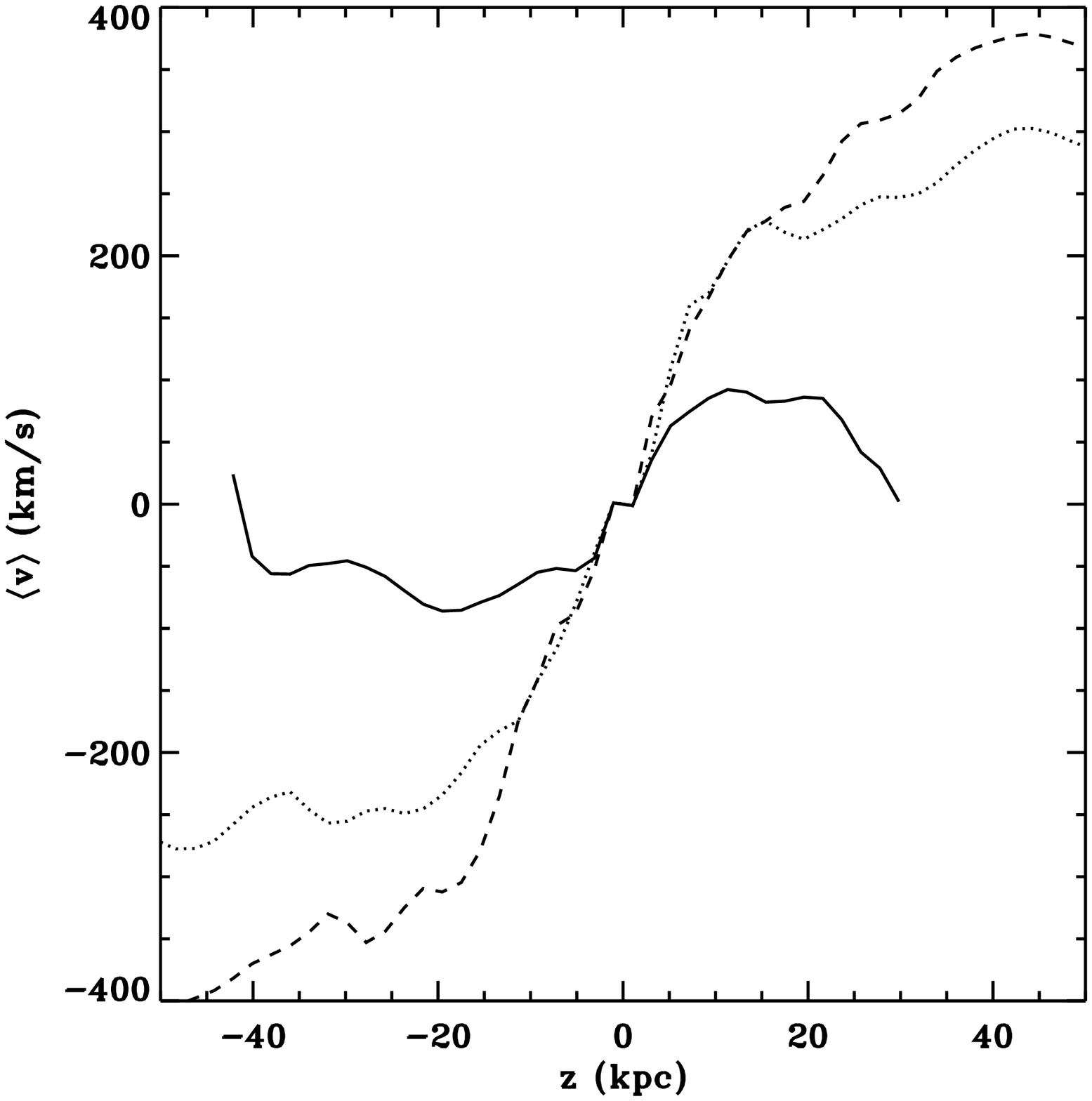}}}
\centering{\resizebox*{!}{7cm}{\includegraphics{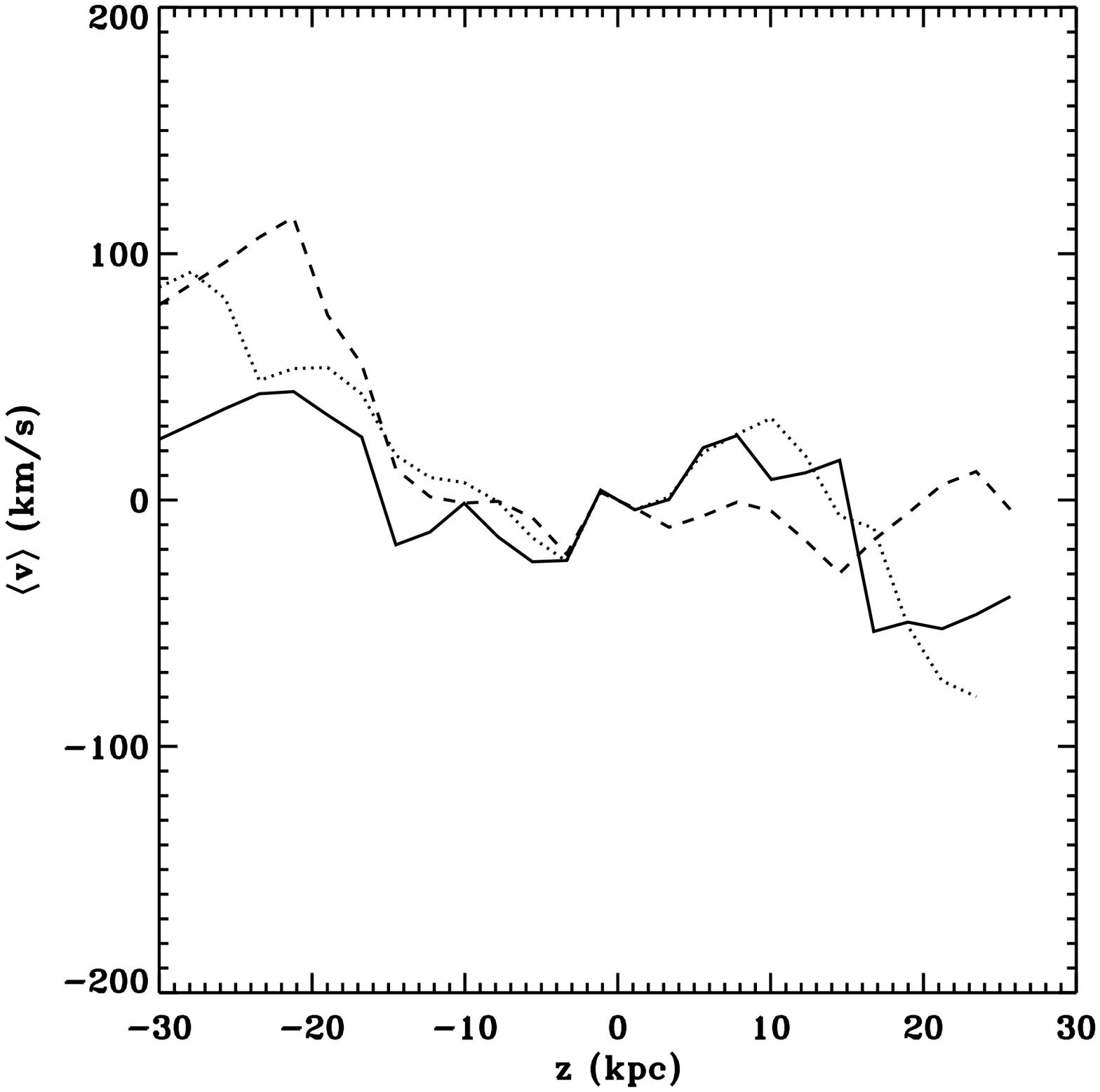}}}
\centering{\resizebox*{!}{7cm}{\includegraphics{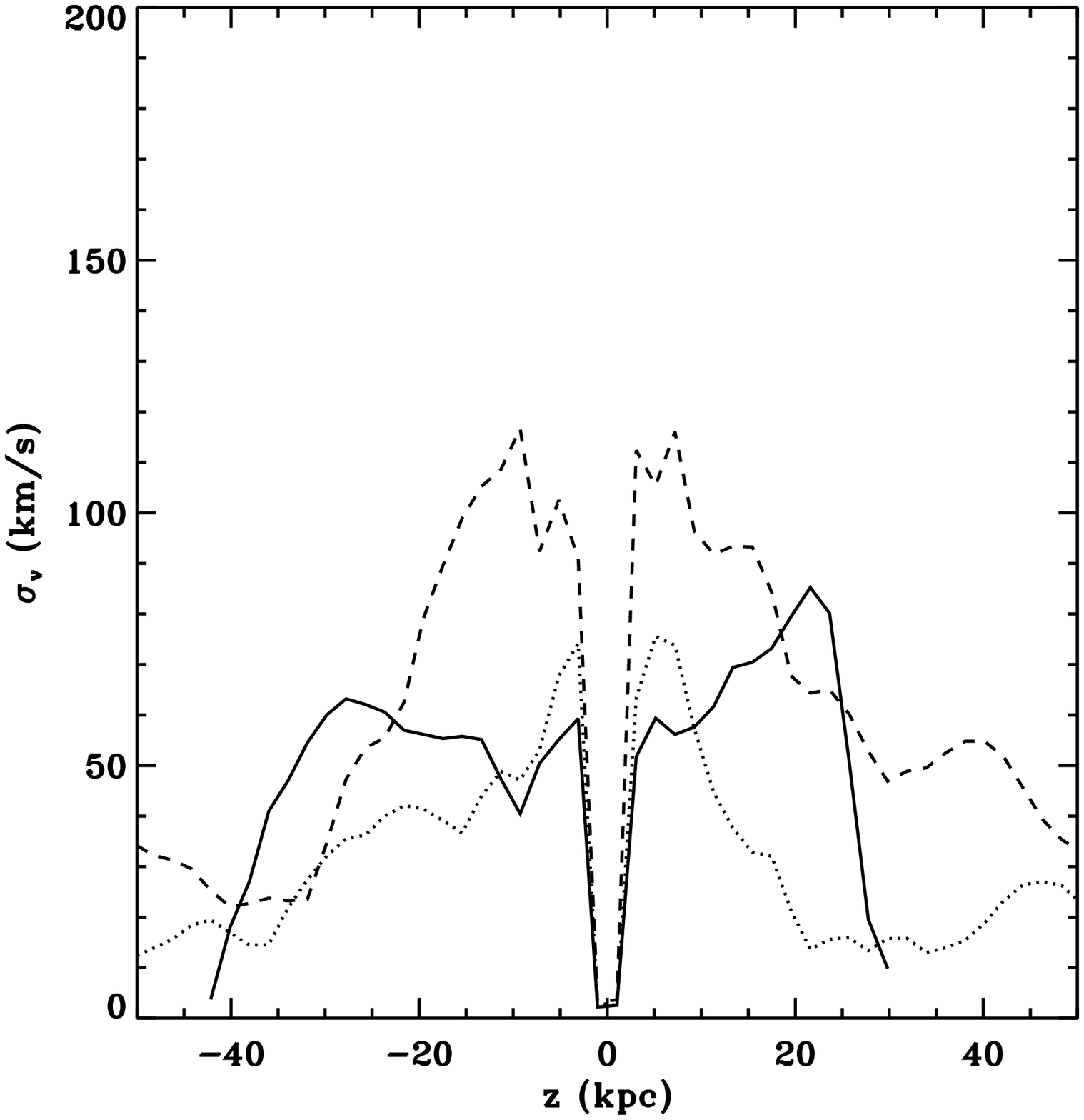}}}
\centering{\resizebox*{!}{7cm}{\includegraphics{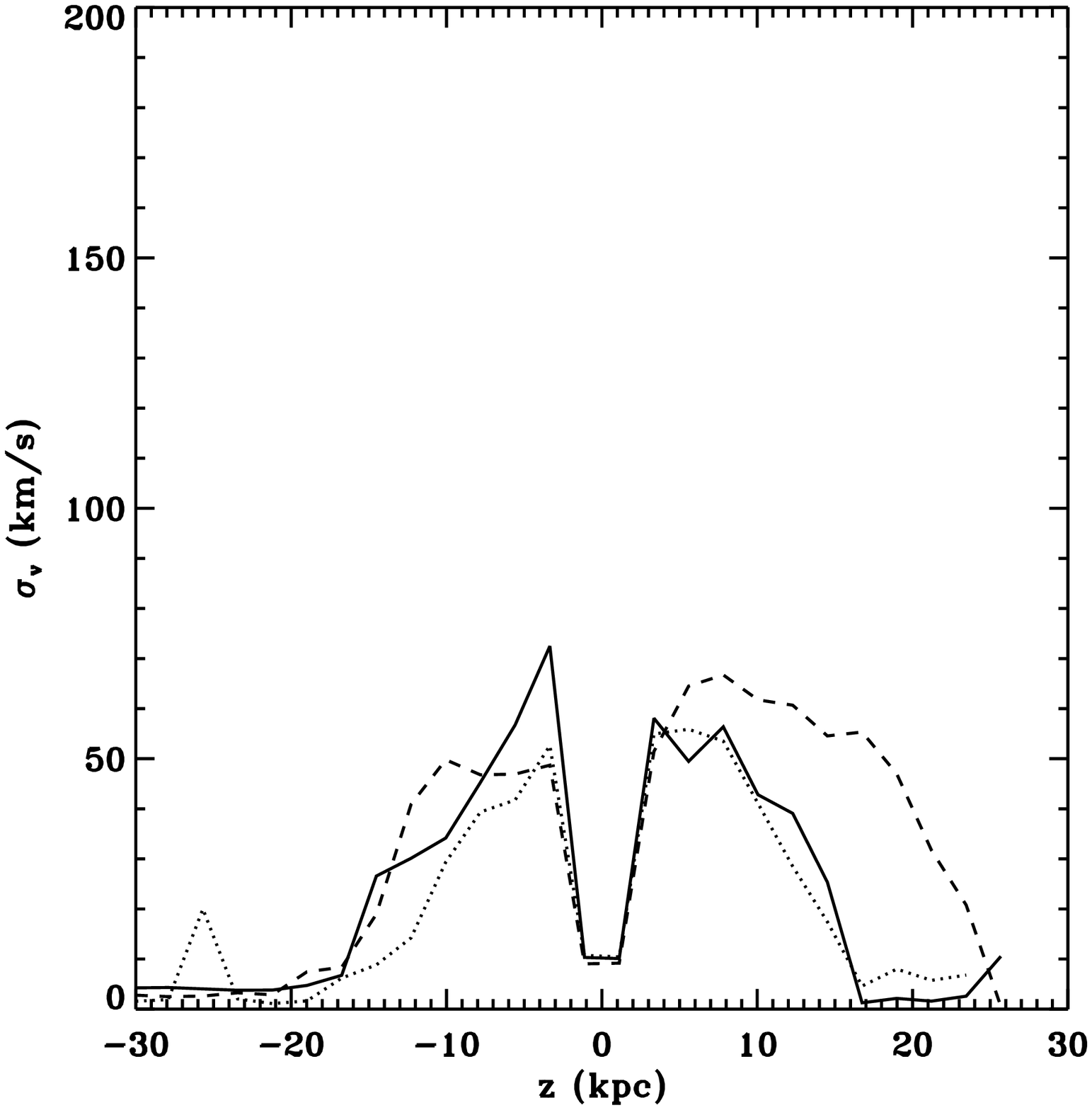}}}
\centering{\resizebox*{!}{7cm}{\includegraphics{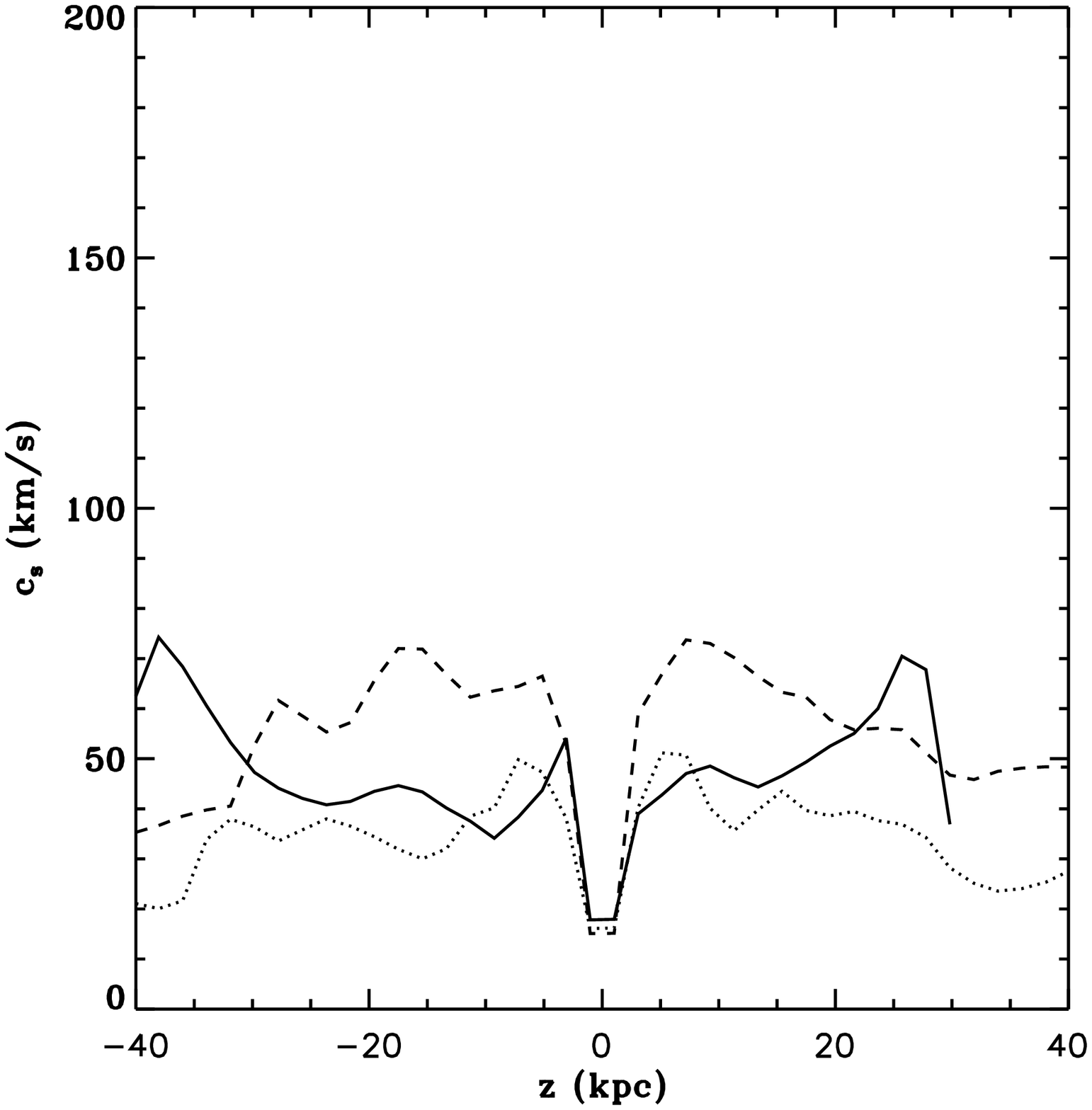}}}
\centering{\resizebox*{!}{7cm}{\includegraphics{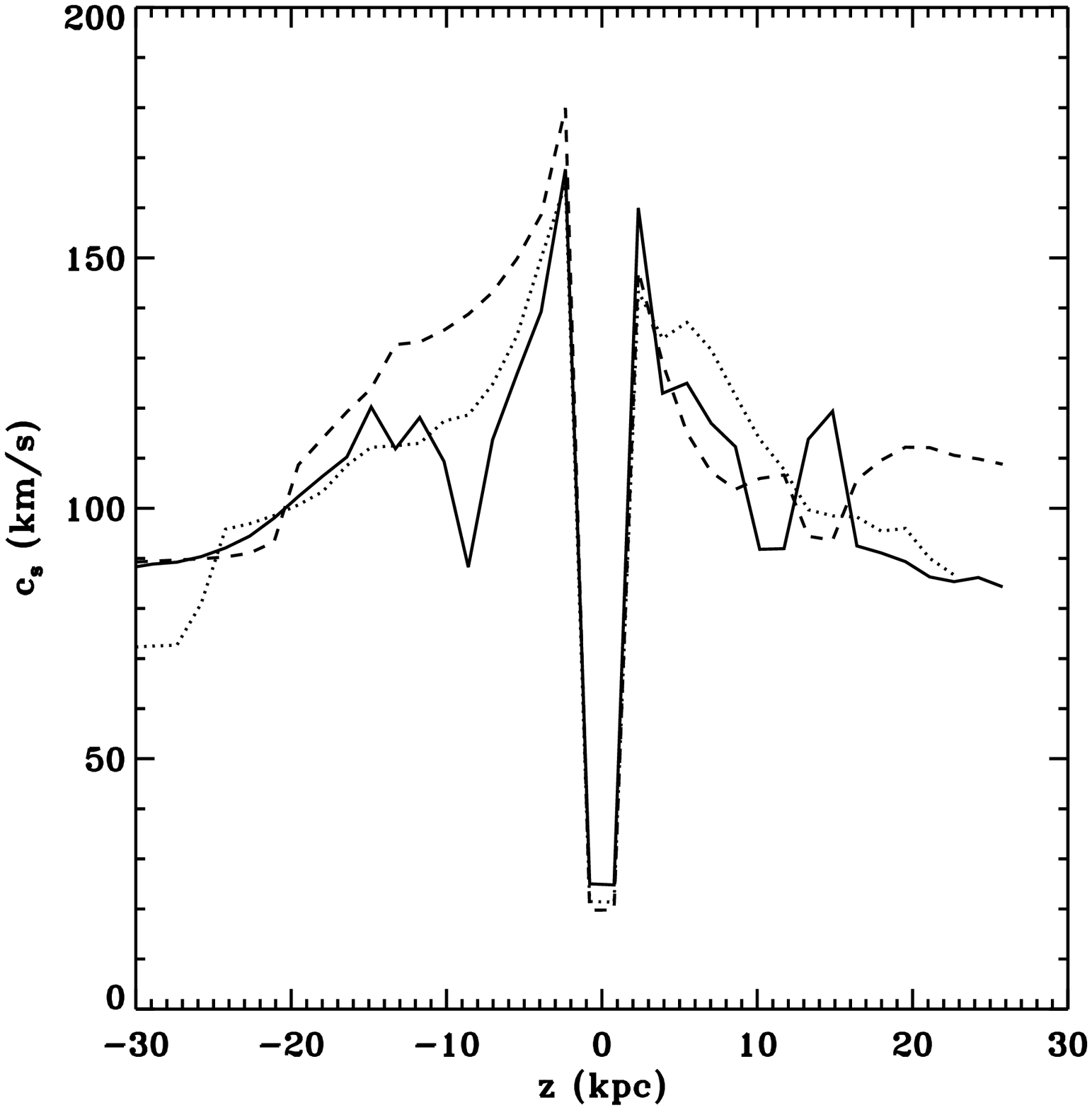}}}
\caption{Mean velocity (up), velocity dispersion (middle), sound speed
(bottom) of  the gas  as a  function of the  projected height  for the
 Sd simulation (left)  and  the Ld simulation (right) at different  times: $t=3 \, \rm  Gyr$ (solid
line),  $t=4.5 \, \rm  Gyr$ (dotted  line), $t=6  \, \rm  Gyr$ (dashed
line).}
\label{vprofiles}
\end{figure*}

Figure~\ref{starsprofiles} shows the  surface density profile of stars
and the corresponding mass--weighted average metallicity as a function
of the  disc radius. In both  cases, we see a  clear exponential disc,
with a bulge--like feature in the  central part and a sharp decline of
surface density in the outer part. We find a bulge radius of roughly 1
kpc and a  disc exponential radius of 2.5 kpc for  the low mass system
and of 4 kpc  for the high mass one. We note  that the outer radius of
the low mass case ($8$ kpc)  in smaller than the rescaled outer radius
of its high--mass counterpart ($10$ kpc). This is due to the different
disc thickness, so that for a given surface density, the threshold for
star formation is reached more easily by the most massive galaxy. This
confirms the  fact that  the polytropic multiphase  model has  a great
importance  in regulating  the gaseous  disc properties  and therefore
star formation efficiencies.  The break in the stellar
surface density corresponds roughly  to the {/it average} radius where
the gas  reaches the density threshold of  star formation ($\rho_0=0.1
\, \rm H .  \rm cm^{-3}$).   Note that the overall amount of stars is
very similar  in both  cases, demonstrating that  the presence  of the
galactic  wind has  no impact  on the  amount of  baryons  locked into
long--lived  stars.   This is  consistent  with  the  rather low  wind
efficiencies we have measured ($\eta_w \simeq 10\%$).  Our simulations
will not solve the  overcooling problem. More importantly, the stellar
metallicity seems also  to be rather insensitive to  the ``wind versus
fountain'' scenario.  We have $Z_* \simeq 0.4 \, \rm Z_{\odot}$ in the
central part of  our $10^{11} \, \rm M_{\odot}$  halo galaxy, while we
have $Z_*  \simeq 0.3 \,  \rm Z_{\odot}$ in  the low mass  case.  Note
that we  have computed a  mass-weighted stellar metallicity so  we are
more sensitive  to the overall  star formation history rather  than to
the  latest OB  stars formed.   Nevertheless,  our wind  model has  no
dramatic impact on the simulated stellar population. although there is
a weak  trend for low mass  galaxies to have  slightly smaller stellar
metallicities than  high mass.  Here again,  as in \cite{dalcanton06},
it is quite difficult to conclude  whether this trend is due to a wind
or to a smaller star formation efficency.

\subsection{Gas kinematics}

The most striking and observable  differences between the wind and the
fountain can  be found in the  gas kinematics in  the halo atmosphere,
below and  above the disc. In order  to mimic as much  as possible the
observational  signatures of  our  simulated galaxies,  as  seen by  a
spectral lines analysis from a distant observer whose line of sight is
aligned with  the disc  rotation axis, we  define the  average metal
velocity as
\begin{equation}
v_{\perp}(z)=\frac{\int_0^{r_d} \rho Z v_z 2\pi r dr }
{\int_0^{r_d} \rho Z 2\pi r dr }~~~{\rm and}~~
v^2_{\perp}(z)=\frac{\int_0^{r_d} \rho Z v^2_z 2\pi r dr }
{\int_0^{r_d} \rho Z 2\pi r dr }
\end{equation} 
The velocity dispersion and the sound speed are defined as usual by
\begin{equation}
\sigma^2_{\perp}(z)=v^2_{\perp}(z)-v_{\perp}(z)^2~~~{\rm and}~~
c^2_s(z)=\frac{\int_0^{r_d} \rho Z c^2_s 2\pi r dr }
{\int_0^{r_d} \rho Z 2\pi r dr }
\end{equation} 
Figure~\ref{vprofiles} shows  these various profiles as  a function of
height for the wind case (left)  and the fountain case (right). In the
former  case, we  clearly see  a  strong symetrical  expansion of  the
metals, with  a strong acceleration up to  200 km/s in 10  kpc, then a
plateau afterwards  when the wind  reaches its terminal  velocity.  In
the latter  case, on the  contrary, we see  a converging flow,  with a
velocity  of  the  order of  100  km/s,  similar  to the  velocity  of
infalling halo  material. At an altitude  of 20 kpc,  this bulk infall
velocity declines and in the same time, the velocity dispersion rises:
we are entering  the galactic fountain for which  the flow is subsonic
and  highly  turbulent. Note  that  in  the  wind case,  the  velocity
dispersion also  rises up to  100 km/s as  we approach the  disc. This
can't  be attributed to  a turbulent  flow, since  it would  have been
sueprsonic (the sound speed is around 50 km/s). It is in fact due to a
geometric  effect: as  soon  as the  wind  expansion velocity  remains
subsonic,  below  an altitude  of  10 kpc,  the  section  of the  wind
increases  with height.  When  the wind  turns  supersonic, around  $z
\simeq 10$ kpc, the section of  the wind is forced to remain constant.
This process is responsible for  the typical noozle--like shape of the
wind.    At  low  altitude,   since  the   wind  is   expanding  quasi
isotropically,  this  mimics  a  supersonic velocity  dispersion  that
vanishes at  higher altitude when the  wind flow is  quasi parallel to
the galaxy rotation axis.

The  $10^{10}  \, \rm  M_{\odot}$  halo  is  characterized by  a  fast
supersonic  flow, expanding  quasi-isotropically at  low  altitude but
quasi-parallel to the rotation  axis at higher altitude.  The $10^{11}
\,  \rm  M_{\odot}$ halo,  on  the  contrary,  is characterized  by  a
converging pattern quickly thermalized close to the disc in a subsonic
turbulent flow.

We applied  the method used  by \cite{prochaskaetal07}
to test whether our simulations  show velocity dispersion in the metal
absorption lines  of QSOs wider than the  velocity dispersion expected
from pure gravitational processes (of the order of the galaxy circular
velocity). In this way, we can  test whether our winds would have been
detected.  In their  paper, \cite{prochaskaetal07} use $\Delta v_{90}$
as the  velocity interval encompassing $90\%$  of the mass  of a given
metal absorber.  We simulate an observer watching the galaxy along the
direction of  the wind  and found  that $90\%$ of  the mass  of metals
corresponds to  metals with velocity  lower than $5\,  \rm km.s^{-1}$.
This is in agreement with our findings that wind efficiencies are very
low in our model. 

\section{Conclusion}

In comparison  with \cite{springel&hernquist03}, we  have very similar
initial conditions  but a lower  wind for $\eta_w$. We  have therefore
more   difficulties  to  solve   the  'over--cooling`   problem.   The
explanation    has    two    reasons     :    on    the    one    hand
\cite{springel&hernquist03}   model  supernovae   explosions   with  a
phenomenological  apporach so  that every  single ejecta  in  the disc
entirely participate  to the large  scale outflow.  In this  work, our
goal was  to model supernovae explosions with  a self-consistent Sedov
blast approach where ejectas are  tighlty coupled to the galactic disc
(injecting energy  in the galactic  turbulent cascade). On  the second
hand,  we  have a  cooling  function  depending  on metallicity:  this
increases     radiative    losses     significantly     compared    to
\cite{springel&hernquist03}.    \cite{kobayashietal06}  performed  SPH
simulations of isolated disks  with star formation and self-consistent
supernovae  feedback, metal  dependent  cooling and  the  same set  of
initial conditions  than ours. They  observed very high  mass ejection
rates for the $10^{10}\, \rm M_{\odot}$ halo with a total mass ejected
rising to 80\% of the total baryons mass.  They noted however a strong
dependance  of  their  results   to  numerical  resolution  and,  more
importantly, to the size of the  feedback radius (it can be as high as
10 kpc  !). They  even observed a  faint wind  in the $10^{12}  \, \rm
M_{\odot}$  halo   case.   \cite{tasker&bryan06}  have   simulated  an
pre-formed isolated  disk with the  AMR code ENZO with  star formation
and self-consistent supernovae feedback  released in pure thermal form
over  one dynamical  time scale.  They  succeed in  shuting down  star
formation but  they didn't  take into account  infalling gas  from the
halo.  \cite{fujitaetal04}  constructed an analytical  prescription to
constrain the mass of the halos that can form large-scale outflow with
starburst  physics.  They  have  performed  several sets  of  2D  grid
simulations of a pre--formed  galaxy with an imposed analytical infall
model  and  found  results very  similar  to  ours  (see Figure  1  in
\citealp{fujitaetal04}) : the key role played by infall and a very low
hydrodynamical efficiency for quiescent modes of star formation.

Using  a  quiescent  model  of star  formation  in  isolated
galaxies, self-consistently simulated from a cooling NFW halo, we have
studied  the  conditions for  a  galactic  wind  to break-out  of  the
ram-pressure exerted by the infalling gas.  Our simulations have shown
that a galactic super wind forms in halos of $10^{10}\, \rm M_{\odot}$
and  no wind  can form  in halo  of mass  greater than  $10^{11}\, \rm
M_{\odot}$,  even for  our most  favorable couple  of  halo parameters
($\lambda=0.04$, $t_0=3 \, \rm Gyr$).  For large galaxies, 
a  galactic fountain appears and expells  metal rich  gas  that cannot
escape from the  galaxy.  Those results are consistent with
our  toy  model  that  predicts  no  wind in  those  galaxies  if  the
hydrodynamical efficiency  is of  a few percents  as suggested  by our
simulations. Using  this rather simple  toy model, we  understand this
failure as due to the ram pressure of infalling material confining the
outflowing wind.

Although galactic  winds develop  in $10^{10}\, \rm  M_{\odot}$ halos,
they are not sufficient to explain the "overcooling" problem. The mass
ejection efficiency obtained in our  simulations $\sim 3-12 \%$ is one
order  of magnitude  lower  than that  observed by~\cite{martin99}  in
starburst galaxies. The  main conclusion of this paper  is that with a
self-consistent   treatment  of   supernovae  feedback   in  idealized
simulations of quiescent star-forming galaxies we cannot reproduce the
observed  mass ejection  rates and  therefore reduce  the  cold baryon
fraction. On the contrary, supernovae feedback is an efficient process
for  metal  enrichment  of  the  IGM.  Cosmological  simulations  with
supernovae feedback  could test this  scenario in a  more quantitative
way.   Using a  more realistic  cosmological setting  may  result into
non--spherical  accretion flows,  and  therefore to  a less  stringent
criterion  for a wind  to break-out.  

Considering  that small  mass  galaxies preferentially
forms along  filaments, non-spherical accretion should  allow the wind
to break-out more easily and  probabily to be more efficient. For more
massive  galaxies, however,  the accretion  of  gas into  the disc  is
likely  to proceed  through a  Virial shock,  and not  along filaments
\citep{dekel&birnboim06}.   Our  conclusions  should therefore  remain
valid,  so that massive  galaxies should  not provide  winds powerfull
enough to enrich the IGM by  SNe feedback (but other way of enrichment
like  gas  stripping  from   satellites  can  be  considered  like  in
\citealp{schindleretal05}).  \cite{dekel&birnboim06}  have computed the
critical mass  between filamentary accretion and Virial  shocks in the
range $10^{11}-10^{12}$ M$_\odot$ depending  on the metallicity of the
infalling  gas.  Combining their  results with  ours, we  can conclude
that  halos  larger than  a  few  $10^{11}$  M$_\odot$ cannot  produce
galactic winds, even within  a realistic cosmological environment.  At
higher redshift,  accretion rates are  believed to be higher  than the
one we have considered here,  so that galactic winds at earlier epochs
are even  more unlikely. A proper  modelling of starburst  (yet to be
invented)  might  also provide  an  easier  route  for increasing  the
feedback efficiency of supernovae-driven outflows.

\section*{ACKNOWLEDGMENTS}
This  work has been  supported by  the Horizon  Project.  Computations
were done at CCRT, the CEA Supercomuting Centre.

\appendix

\section{Numerical scheme for Supernovae}
\label{sn_scheme}

We define the mass vanished in the star formation process as :
\begin{equation}
\left ( \Delta m_g \right )_{SF}=m_*(1+\eta_{SN} +\eta_W) \, ,
\label{dmgstar}
\end{equation}
where the second  term of the right part is equal  to the total debris
mass $m_d$. We  have also introduce the mass  loading factor $\eta_W$,
the  parameter that  determines the  gas mass  carried in  debris. And
$m_*$ is equal to
\begin{equation}
m_*={ \rho_0 \delta x^3 \over 1+ \eta_{SN}+\eta_W} \, ,
\end{equation}
$\delta x^3$  is the  volume of  the cell where  the star  particle is
spawned. Then if we introduce the fraction of debris particles :
\begin{equation}
f_d={\eta_{SN} +\eta_W \over 1+ \eta_{SN} +\eta_W} \, , 
\end{equation}
we  can simplify equation  (\ref{dmgstar}) by  saying that  a fraction
$f_d$ of the  gas consumed goes into debris  particles, and a fraction
$1-f_d$ goes into star particle.

The maximum speed of a debris is given by
\begin{equation}
u_d={ u_{SN} \over \sqrt {1+ \eta_W/\eta_{SN}}} \, .
\end{equation}
$u_{SN}$ is  the typical velocity corresponding to  the kinetic energy
released in one single  supernova explosion ($u_{SN}\simeq 3200 \, \rm
km.s^{-1}$).

Thus, the energy released to the gas by the debris is
\begin{equation}
E_d=\eta_{SN} {m_* \over M_{SN}} E_{SN} \, .
\end{equation}
$M_{SN}$ and $E_{SN}$ are respectively the typical progenitor mass and
energy  of an  exploding type  II  supernova (i.  e. $M_{SN}=10\,  \rm
M_{\odot}$  and  $E_{SN}=  10^{51}   \,  \rm  erg$).   The  energy  is
independent from the mass loading factor. Then, we have to compute the
moment conservation  for constraining this free parameter.  We want to
reproduce a Sedov blast wave, and  so the velocity of the gas when the
debris  are  coupled  with  it   is  the  Sedov  speed  of  the  shock
propagation.
\begin{equation}
u_{Sedov}=\beta {2 \over 5} \left ( { {E_d} \over \rho_0 \Delta x^3} \right )^{1/2} \, ,
\end{equation}
$\beta$  is a multiplication  factor nearly  equal 1,  we admit  it is
equal to  1, $\rho_0$ is  the density of  the gas where  the explosion
takes place and $\Delta x$ is  the radius of the shock form the center
of the explosion. We can also express $u_{Sedov}$ in terms of $u_{SN}$
:
\begin{equation}
u_{Sedov}={\sqrt{2} \over 5} \left [ \eta_{SN} \left ( {\delta x \over \Delta x } \right ) ^3 {1 \over 1+\eta_{SN}+\eta_W} \right ] ^{1/2} u_{SN}
\end{equation}

The Sedov  blast wave is only  valid in an homogeneous  fluid, then it
seems surprising  to apply it in  regions of star  formation where the
gas condenses in high overdensity regions of molecular clouds. That is
the reality, but our simulations do not reproduce such high resolution
features, and  the assessment  on the polytropic  trend of the  gas at
high  densities implies  more pressurized  regions in  the  disc. This
behavior at high densities transform high over-density regions of star
formation  into smoother  regions  at the  scale  of one  cell of  the
numerical grid.  That is the  reason that allows  us to use  the Sedov
blast wave model.  We want the gas to have the moment of a Sedov blast
wave, so the conservation of the moment gives us :
\begin{equation}
m_d u_d=(m_g+m_d) \bar u_{Sedov} \, ,
\end{equation}
where  $m_g=4/3 \pi  \rho_0  \Delta x^3$  is  the gas  mass where  the
supernovae explosion occur, and $\bar u_{Sedov} \simeq u_{Sedov}/3$ is
the mean velocity of the gas carried along the shock.

As we know all the quantities of this last equation, we can constraint
our free parameter $f_d$, and obtain
\begin{equation}
\left [ 1+ {1 \over f_d} {4 \over 3} \pi \left ( {\Delta x \over \delta x }\right )^3 \right ] {\sqrt{2}\over15} \left [ f_d \left ( {\delta x \over \Delta x} \right )^3 \right ]^{1/2}=1 \, .
\end{equation}
If we  want that our  debris propagates to  the nearest cells,  let us
assume that $ \Delta x \simeq  1.5$, and we find the optimal value for
$f_d \simeq 0.5$. Thus for  a typical value of $\eta_{SN}=0.1$ we find
for the mass  loading factor, our only free  parameter, $\eta_W \simeq
1$.

\bibliographystyle{aa}
\bibliography{article}

\end{document}